\documentstyle[aps,epsf]{revtex}
\def\`#1{\if#1i{\accent18 \i}\else{\accent18 #1}\fi}
\def\'#1{\if#1i{\accent19 \i}\else{\accent19 #1}\fi}
\def\gcmsq{${\rm g}/{\rm cm}^2$}
\def\GeV2{${\rm GeV}^2$}

\newcommand{\AmS}{{\protect\the\textfont2
  A\kern-.1667em\lower.5ex\hbox{M}\kern-.125emS}}

\begin{document}
\draft
\date{March 22, 2001}

\title{Extended Air Showers and Muon Interactions.
}
\author{A.~ N.~Cillis \cite{a} and S.~J.~Sciutto \cite{b}}
\address{Laboratorio de F\'isica Te\'orica\\
Departamento de F\'isica\\
Universidad Nacional de La Plata\\
C. C. 67 - 1900 La Plata\\
Argentina}

\maketitle
\begin{abstract}
The objective of this work is to report on  the influence of muon
interactions on the development of air showers initiated by
astroparticles. We make a comparative study of the different
theoretical approaches to muon bremsstrahlung and muonic pair
production interactions.
A detailed algorithm that includes all the relevant characteristics of such
processes has been implemented in the AIRES air
shower simulation system. We have simulated ultra high
energy showers in different conditions in order to measure the
influence of these  muonic electromagnetic interactions. We have found
that  during the
late stages of the shower development (well beyond the shower
maximum) many global observables are significantly modified in relative
terms when the mentioned interactions are taken into account. This is
most  evident in the case of the electromagnetic component of very inclined
showers. On the other hand, our simulations indicate that the studied processes
do not induce significant changes either in the position of the shower
maximum or the structure of the shower front surface.
\end{abstract}


\section{Introduction}

When an ultra high energy astroparticle interacts with an atom of the Earth's
atmosphere, it produces a shower of secondary particles that continues
interacting and generating more secondary
particles. The study of the characteristics of air showers initiated by
ultra high energy cosmic rays is of central importance. This is due to
the fact that in our days such primary particles cannot be detected
directly; instead, they must be studied from different measurements of the  air
showers they produce.

We have been studying the physics of air showers for several years.
We started working on the topic of the electromagnetic processes in
air showers analyzing the modifications in the
shower development  due to the  reduction of
the electron bremsstrahlung and electron pair production by the
Landau-Pomeranchuk-Migdal effect and the dielectric suppression
\cite{PRD}; and we also studied the influence of the geomagnetic field in
an air shower \cite{GF}. 

The main goal of this work is to analyze other radiative processes that take
place during the development of an ultra high energy air shower. We
have  studied the processes of muon bremsstrahlung,  muonic pair
production (electron and  positron) and muon-nucleus interaction.  At
high energies these processes become important and dominate the
energy  losses of the energetic muons that are present in an air
showers.  The mentioned mechanisms  are
characterized by small cross sections, hard spectra, large energy
fluctuations  and generation of electromagnetic sub-showers for the
case of muon bremsstrahlung and muonic pair production, and hadronic sub-showers
for the case of muon-nucleus interaction. As a consequence,  the
treatment of such energy losses as uniform and continuous processes is for
many purposes inadequate \cite{PDG}.

We have studied the three mentioned processes concluding that the muon
nucleus interaction has less probability than the other ones
to  produce hard events, and therefore, to generate sub-showers and 
introduce  significant modifications in the air shower development. 
For this  reason, this work will be primarily focused on studying the
consequences of the purely electromagnetic  processes, namely, muon
bremsstrahlung and muonic pair production.

In order to analyze the influence  of muon bremsstrahlung and  muonic
pair  production on the air shower observables, we have  developed new
procedures for these mechanisms and have incorporated
them in AIRES (AIRshower Extended Simulations) \cite{Aires,Aires2}.
Using  the data generated with the AIRES code, we have studied the changes
introduced by those processes on the different physical quantities.
  
This work is organized as follows: in section II we briefly  review
the  theory of muon bremsstrahlung, muonic pair
production and muon-nucleus interaction. At the end of this section we
compare the three effects and analyze under which conditions they can
modify  the development of air showers. In
section III we show the results of our simulations. 
Finally  we present our conclusions and comments in section IV.

\section{Theory }

\subsection{Theory of Muon Bremsstrahlung} 

The first approach to the muon bremsstrahlung (MBR) theory  was due to
Bethe and Heitler \cite{BH,BH2,Rossi}.
Their results can be reproduced by the standard method of QED
\cite{Greiner}  similarly as in the case of  electron bremsstrahlung. 
Bethe and Heitler also considered in their calculation the screening
of the atomic electrons.

After this first formulation some  corrections were introduced. Kelner,
Kokoulin and Petrukhin \cite{KKP} also considered the interactions 
with the atomic electrons. 
The nuclear form factor was investigated by Christy and Kusaka
\cite{CK} for the first time and then by Erlykin \cite{E}. Petrukhin
and Shestakov \cite{PSH} found that the influence of the nuclear
form factor is more important than the predictions of  previous
papers. These last
results have been confirmed by  Andreev et al. \cite{ABB}
who also considered the excitation of the nucleus. 
In the following paragraphs we give some details of these different
approaches for MBR theory.

\subsubsection{MBR with the effect of the screening by the atomic electrons}
It is possible to  reproduce the results found by Bethe and Heitler
\cite{BH,BH2,Rossi} (in the case of no screening) performing the
calculations of the  Feynman diagrams (see figure \ref{fig:fdmb}) at
first order of perturbation theory.
For energies that are large compared with the muon mass,  the MBR differential
cross section integrated over final muon and photon angles
takes the form:
\begin{equation}
\sigma(v,E)dv=\alpha\left[\frac{2Zr_0m_e}{m_\mu}\right]^2\left[(2-v+v^2)
-\frac{2}{3}(1-v)\right]\frac{dv}{v}
\label{sigmab}
\end{equation}
where $E$ is the primary energy of the muon, $k$  is the photon energy,
$v=k/E$  is the fraction energy transferred to the photon, $r_0$ is
the classical electron radius ($r_0=2.81794092\times 10^{-15}$ m), $m_e$
($m_\mu$) is the electron (muon) mass and $\alpha$ is the fine
structure constant ($\hbar=c=1$ throughout this paper).

When the atomic state involved is not changed, the effect of atomic electrons
(screening) is taken into account by introducing the elastic atomic
form factor in the cross section for bremsstrahlung under the effect
of a Coulomb center \cite{Tsai}.

As we have just mentioned before, Bethe and Heitler \cite{BH},
took into account in their calculation the influence of the screening. 
The  atomic  electrons change the Coulomb potential of the nucleus in
the following way \cite{BH,Tsai}:
\begin{equation}
V=\frac{1}{q^2}\left[Z-F(q)\right]^2
\label{V}
\end{equation}
where $Z$ is the charge of the nucleus, $q$ is the momentum transferred
to the nucleus and $F$ is the atomic form factor, that is (in
spherical coordinates):
\begin{equation}
F(q)=\int\rho(r)e^{iqr}d\Omega
\label{F}
\end{equation}
where $\rho(r)$ is the density of the atomic electrons at the distance
$r$ of the nucleus. Bethe and Heitler assume the Fermi distribution
for this density that is,
\begin{equation}
\rho(r)=\frac{\rho_0}{1 + exp\left[\frac{r-\alpha}{\beta}\right]}
\label{rho}
\end{equation}
where the constants $\alpha$ and $\beta$ are different for each of the
elements. This distribution describes adequately all elements with
$Z\geq10$.   
The Fermi radius of the atom  is given by:
\begin{equation}
a(Z) = a_0Z^{-1/3}
\label{a}
\end{equation}
where $a_0$ is the Bohr radius.\\
The screening effect becomes important when $F$ in equation (\ref{V})
is comparable with $Z$. This occurs when $q$ is of the order (or
smaller than) the reciprocal atomic radius,  that is:
\begin{equation}
q\ll\frac{Z^{1/3}}{a_0}=\alpha m_e Z^{1/3}
\label{q}
\end{equation}
In this case the phase ($q r$) in equation (\ref{F}), and
thus  $F(q)$, are small.

Due to the fact that the differential cross section is proportional
to $1/q^2$, the largest contribution to the radiation cross section originates
from the region where the momentum transferred, $q$, is small. Let 
$q_{min}$ be the minimum of $q$. Using equation (\ref{q}), the
condition for the screening to be effective reads 
\begin{equation}
q_{min}\leq q\ll \alpha m_e Z^{1/3}
\label{qbis}
\end{equation}
The  minimum value of $q$ occurs when  the momentum of the
muon is parallel to the emitted photon, 
\begin{equation}
q_{min}=\delta=p_1-p_2-p_k
\label{qmin}
\end{equation}
where $p_1$ ($p_2$) is the initial (final) momentum of the muon and
$p_k$ is the  momentum of the emitted photon. When the energies
considered are larger compared with the muon mass, the last equation
reduces to
\begin{equation}
\delta=\frac{m_\mu^2}{2E}\, \frac{v}{1-v},
\label{delta}
\end{equation}
and from equations (\ref{qbis}), (\ref{qmin}) and (\ref{delta}) we can
write
\begin{equation}
\frac{E(E-k)}{k}\gg \frac{m_{\mu}^2Z^{-1/3}}{2m_e\alpha}
\label{E}
\end{equation}
It is common to use the ratio between the atomic shell radius (\ref{a}) and
the distance from the nucleus R, as a parameter that gives a
quantitative estimation of the importance of the screening effect:
\begin{equation}
\gamma=\frac{a(Z)}{R}.
\label{gamma}
\end{equation}
One can then estimate the distance from the nucleus using the uncertainty
relation ($R=1/\delta$) and  equation (\ref{a}), so $\gamma$ can be
written as 
\begin{equation}
\gamma=\frac{1}{\alpha m_e Z^{1/3}}\frac{m_{\mu}^2}{2E}\,\frac{v}{1-v}
\label{gamma2}
\end{equation}
The limits $\gamma\rightarrow 0$ and $\gamma \gg 1$ correspond
respectively  to the cases of appreciable and negligible screening effect.
 
When the influence of the atomic electrons is taken into account, the
integration over angles in the differential cross section of muon
bremsstrahlung \cite{BH} can only be carried out numerically.
Accordingly to the calculation of Bethe and Heitler  \cite{BH2} the
MBR  cross section can be written in the following way (for $E \gg m_\mu$)
\begin{equation}
\sigma(v,E)dv=\alpha\left[\frac{2Zr_0m_e}{m_\mu}\right]^2\left[(2-2v+v^2)\phi_1(\delta)-\frac{2}{3}(1-v)\phi_2(\delta)\right]\frac{dv}{v}
\label{sigmab1}
\end{equation}
$\phi_1(\delta)$ and $\phi_2(\delta)$ are the well known functions
displayed in figure 1 of  Bethe and Heitler's paper \cite{BH2,Rossi}.
Since the relative difference
$|\phi_1(\delta)-\phi_2(\delta)|/\phi_1(\delta)$ remains less than 3\%
for all $\delta$, the approximation
$\phi_1(\delta)=\phi_2(\delta)\equiv\phi(\delta)$ is justified and
equation  (\ref{sigmab1}) can be put in the following way:
\begin{equation}
\sigma(v,E)dv=\alpha\left[\frac{2Zr_0m_e}{m_\mu}\right]^2\left[\left(\frac{4}{3}-\frac{4}{3}v+v^2\right)\phi(\delta)\right]\frac{dv}{v}
\label{sigmab2}
\end{equation}
Petruhkin and Shestakov \cite{PSH} give an analytical expression for
$\phi(\delta)$ for any degree of screening. Their result can be
expressed as
\begin{equation}
\phi(\delta)=\ln\left\{\frac{189m_{\mu}}{m_e}Z^{-\frac{1}{3}}\left[1+\frac{189\sqrt{e}}{m_e}\delta Z^{-\frac{1}{3}}\right]^{-1}\right\}
\label{phi}
\end{equation}
This $\phi(\delta)$ expression  differs from the original $\phi_1$ and
$\phi_2$  of Bethe and Heitler theory less than 3.3 \% for $\delta=0$
and  about 10 \% when $\delta=2\alpha me Z^{\frac{1}{3}}$ (this
corresponds to $\gamma =2$).
 
\subsubsection{Correction due to MBR by the atomic electrons.}
Another contribution that must be taken into account is the MBR with
the  atomic electrons \cite{KKP}. This correction
is especially important for light nuclei.

It has been found \cite{KKP} that the usual transformation $Z^2$ replaced by $Z(Z+1)$ in the
differential cross section  is not accurate enough to take into
account adequately  the influence of the above mentioned process.
Instead, the inelastic atomic form factor must be included \cite{Tsai} in order
to  take into account the electron binding within the atom. 
If $d\sigma_0(E,k,q)$ is the differential cross section for MBR by free
electrons then the MBR by the atomic electrons is given by \cite{Tsai}:
\begin{equation}
d\sigma(E,k) = Z \int_0^{q_{max}}F_{a}^{in}(q)d\sigma(E,k,q)
\label{sigmaelec}
\end{equation}
where $F_a^{in}(q)$ is the inelastic atomic form factor.

In the work of  Kelner et. al. \cite{KKP}  the inelastic form factor
was calculated according to the Thomas-Fermi model \cite{Tsai}. Those
authors have found the following formula that approximates
the differential MBR cross section for  muon scattering by atomic
electrons \cite{KKP}
\begin{equation}
\sigma(v,E)dv=\alpha\left[\frac{2Zr_0m_e}{m_\mu}\right]^2\left[\left(\frac{4}{3}-\frac{4}{3}v+v^2\right)\phi_e(\delta)\right]\frac{dv}{v}
\label{sigmab3}
\end{equation}
where
\begin{equation}
\phi_e(\delta)= \ln\left[\frac{m_{\mu}/\delta}{m_{\mu}\delta/m_e^2
+\sqrt{e}}\right] -\ln\left[1+\frac{m_e}{\delta B'Z^{-2/3}\sqrt{e}}\right],
\label{phie}
\end{equation}
and $e=2.718$, $B'=1429$. 

\subsubsection{Correction due to nucleus form factor and nucleus excitation.}

The influence due to the nuclear form factor is usually taken into
account as a correction to $\phi(\delta)$. Petrukhin and
Shestakov \cite{PSH} have found (in the case of nucleus with $Z>10$) that
the modification due to the nuclear size can be accounted for by  changing
the  equation corresponding to point nucleus, that is equation 
(\ref{phi}), by
\begin{equation}
\phi(\delta)=\ln\left\{\frac{189m_{\mu}}{k m_e}Z^{-\frac{2}{3}}\left[1+\frac{189\sqrt{e}}{m_e}\delta Z^{-\frac{1}{3}}\right]^{-1}\right\}
\label{phibis}
\end{equation}
where $k=3/2$.

Petrukhin and Shestakov \cite{PSH} proved that the inclusion of the
elastic nuclear form factor decreases appreciably the differential
bremsstrahlung cross section (by approximately by 10-15 \% when $Z>10$). This
result is in  agreement with the one of Andreev et. al. \cite{ABB}.

It is also possible to take into account an additional correction due
to the  nuclear level  excitations, but this contribution  amounts to
only about 1 \% of the elastic one in the case of nuclei with $Z\simeq
10$ \cite{ABB}.

\subsubsection{Final result taking into account the different
corrections analyzed above.}

To define the algorithms to be used in our calculations we have considered
the  correction due to atomic screening, the MBR by atomic electrons
and  the nuclear form factor.  
The final result for the MBR differential cross section can be written
as :
\begin{equation}
\sigma(v,E)dv=\alpha\left[\frac{2r_0m_e}{m_\mu}\right]^2
Z(Z\phi+\phi_e) \left[\frac{4}{3}-\frac{4}{3}v + v^2 \right]\frac{dv}{v}
\label{sigmab4}
\end{equation}
where  $\phi$ is given by equation (\ref{phibis}) and $\phi_e$ is
given by  equation (\ref{phie}).

The expression for the differential cross
section diverges when  the photon energy tends to
zero (infrared divergence).  In order to overcome this mathematical problem,  it is usual
to put a cut off in the
photon  energy, $k_c$. Therefore, the total cross section for a photon
emitted with energy bigger than $k_c$ is calculated by
\begin{equation}
\sigma(Z,T,k_c)=\int_{k_c}^{T}\frac{d\sigma(Z,T,k)}{dk}dk
\label{sigmatotalb}
\end{equation}
Below the cut off  $k_c$, the mean energy loss by the muon
due to bremsstrahlung is given by
\begin{equation}
E_{loss}^{brem}(Z,T,k_c)=\int_{0}^{k_c}k\frac{d\sigma(Z,T,k)}{dk}dk
\label{Elossb}
\end{equation}
It is worthwhile mentioning that, as in the case of electron
bremsstrahlung,  the cut off in the photon energy
can be  naturally introduced if the Landau-Pomeranchuk-Migdal effect and
dielectric suppression \cite{PRD} are taken into account \cite{Politiko,CS}.

\subsection{Theory of Muonic  Pair Production}

In the lowest significant order of perturbation theory, the muonic pair
production (MPP) is a 4th order process in QED. Two types of
diagrams are present, respectively labelled $\mu$ and $e$ in figure
\ref{fig:fdmpp}. The main contribution to the total cross section
and to the energy loss of muons comes from  the $e$-diagrams.
The $\mu$-diagrams  have to be taken into account if the energy
fraction transferred is large \cite{KP}.

Similarly as in the case of MBR, there are several corrections that
need to be taken into account in the MPP calculations. Such
corrections are of the same kind that the ones introduced for MBR. For
brevity we are not going to review them in full detail. Instead, we
are going to describe the final results which include all the
relevant corrections. 

Racah \cite{R} calculated for the first time the MPP cross section in
the relativistic region, but without taking into account the atomic and the
nuclear form factor.  Thereafter, Kelner \cite{K}  included the
correction due to the screening of the atomic electrons.
The analytical expression for any degree of screening was introduced
by Kokoulin and Petrukhin \cite{KP}.
Those authors also take into account  \cite{KP2} the correction due
to the nuclear form factor. We wish to emphasize that the
influence of the nuclear size is more important when  the energy
transferred to the pair is large \cite{KP2}. This last case is
important for ultra high energy air showers and therefore the nuclear
size  effect needs to be included in the algorithms used for the simulations. 

If the atomic and nuclear form factors are taken into account, the MPP
differential cross section can be expressed as \cite{KP2}:
\begin{equation}
\frac{d^2\sigma}{d\nu d \rho}=\frac{2}{3\pi}Z(Z+1)(\alpha
r_0)^2\,\left(\frac{1-\nu}{\nu}\right)\left[\phi_e + 
(\frac{m_e}{m_{\mu}})^2 \phi_{\mu} \right]
\label{pp}
\end{equation}
where
\begin{equation}
\nu=\frac{E^{+} +E^{-}}{E},
\label{nu}
\end{equation}
\begin{equation}
\rho=\frac{E^{+} - E^{-}}{E^{+}+E^{-}},
\label{rho2}
\end{equation}
$E^{\pm}$ is the total energy of the $e^{\pm}$,  and
\begin{equation} 
\phi_{e,\mu}=B_{e,\mu}L_{e,\mu},
\label{phiemu}
\end{equation}
with
\begin{equation}
B_e=
\left[(2+\rho^2)(1+\beta)+\epsilon(3+\rho^2)\right]
        \ln\!\left(1+\frac{1}{\epsilon}\right)
+\frac{1-\rho^2-\beta}{1+\epsilon}-(3+\rho^2),
\label{Be}
\end{equation}
\begin{equation}
B_{\mu}=
\left[(1+\rho^2)(1+\frac{3\beta}{2})-\frac{1}{\epsilon}(1+2\beta)(1-\rho^2)\right]\ln(1+\epsilon)+\frac{\epsilon(1-\rho^2-\beta)}{1+\epsilon}-(1+2\beta)(1-\rho^2),
\label{Bmu}
\end{equation}
\begin{equation}
L_e=\ln\left[\frac{C
Z^{-1/3}\sqrt{(1+\epsilon)(1+Y_e)}}{1+\frac{2m_e\sqrt{e}C Z^{-1/3}(1+\epsilon)(1+Y_e)}{E\nu(1-\rho^2)}}\right]-\frac{1}{2}\ln\left[1+\left(\frac{3m_eZ^{1/3}}{2m_{\mu}}\right)^2(1+\epsilon)(1+Y_e)\right],
\label{Le}
\end{equation}
\begin{equation}
L_{\mu}=\ln\left[\frac{\frac{2}{3}\frac{m_{\mu}}{m_e}C Z^{-2/3}}{1+\frac{2m_e\sqrt{e}C Z^{-1/3}(1+\epsilon)(1+Y_{\mu})}{E\nu(1-\rho^2)}}\right],
\label{Lmu}
\end{equation}
\begin{equation}
Y_e = \frac{5-\rho^2
+4\beta(1+\rho^2)}{2(1+3\beta)\ln(3+1/\epsilon)-\rho^2-2\beta(2-\rho^2)},
\label{Ye}
\end{equation}
\begin{equation}
Y_{\mu} = \frac{4+\rho^2
+3\beta(1+\rho^2)}{(1+\rho^2)(\frac{3}{2}+2\beta)\ln(3+\epsilon)+1-\frac{3}{2}
\rho^2},
\label{Ymu}
\end{equation}
\begin{equation}
\beta=\frac{\nu^2}{2(1-\nu)} ,
\label{beta}
\end{equation}
\begin{equation}
\epsilon=\left(\frac{m_{\mu} \nu}{2m_e}\right)^2  \frac{1-\rho^2}{1-\nu}.
\label{epsilon}
\end{equation}
C is equal to 189, and  the kinematic ranges
of $\nu$ and $\rho$ are:
\begin{equation}
\frac{4m_e}{E}=\nu_{min}\leq\nu\leq\nu_{max}=1-0.75\sqrt{e}\frac{m_{\mu}}{E}Z^{-\frac{1}{3}}
\label{nu2}
\end{equation}
\begin{equation}
0=\rho_{min}\leq\rho(\nu)\leq\rho_{max}=\left[1-\frac{6m_{\mu}^2}{E^2(1-\nu)}\right]\sqrt{1-\frac{\nu_{min}}{\nu}}
\label{rho3}
\end{equation}
The total cross section for the emission of an $e^{+}e^{-}$ pair is:
\begin{equation}
\sigma(Z,T,E_c)=2\int_{\nu_c}^{\nu_{max}}d\nu\int_{0}^{\rho_{max}(\nu)}
d\rho \frac{d^2\sigma}{d\nu d \rho}
\label{sigmatotalpp}
\end{equation}
where $E_c$ is the energy cut-off that has to be introduced to
overcome the infrared divergence of equation (\ref{pp}). 
Below the energy $E_c$ the process can be treated as a continuous
energy loss. The mean value of the energy lost by the incident muon
due to $e^{+}e^{-}$ pair production  with energy below $E_c$ is
\begin{equation}
E_{loss}^{pair}(Z,T,E_c)= 2E \int_{\nu_{min}}^{\nu_c} \nu d\nu \int_{0}^
{\rho_{max(\nu)}}d\rho \frac{d^2\sigma}{d\nu d \rho}
\label{Elosspp}
\end{equation}
\subsection{Muon-nucleus interaction}

The nuclear interaction of high energy muons is theoretically much
less understood than the purely electromagnetic processes studied in
sections A and B. 

Borog and Petrukhin  \cite{BP} calculated a formula for the differential cross
section of this process based on  Hand's Formalism \cite{Hand} for inelastic
muon scattering, and semi-phenomenological inelastic form factor;  and
includes nuclear shadowing effect. Their final result is given by:
\begin{equation}
d\sigma(E, k)=\Psi(k)\Phi(E,v)
\label{sigma3}
\end{equation}
where
\begin{equation}
\Psi(k)=\frac{\alpha}{\pi}A_{\rm eff}\sigma_{\gamma N}(k)\frac{1}{k},
\label{Psi}
\end{equation}
\begin{equation}
\Phi(E,v)=v-1 +
\left[1-v+\frac{v^2}{2}\left(1+\frac{2\mu^2}{\Lambda^2}\right)\right]\ln\left[\frac{\frac{E^2(1-v)}{\mu^2}\left[1+\frac{\mu^2v^2}{\Lambda^2(1-v)}\right]}{1+
\frac{Ev}{\Lambda}\left[1+\frac{\Lambda}{2M}+\frac{Ev}{\Lambda}\right]}\right], 
\label{Phi}
\end{equation}
$k$ is the energy lost by the muon, $v=k/E$, $m_{\mu}$ ($M$) is
the muon (proton) mass and $\Lambda^2=0.4$ \GeV2. The nuclear shadowing
effect is taken into account in  $A_{\rm eff}$ according to the
parameterization of Brodsky \cite{Brod}
\begin{equation}
A_{\rm eff}=0.22A+ 0.87A^{0.89},
\label{Aeff}
\end{equation}
where $A$ is the atomic mass. The photo nuclear cross section,
$\sigma_{\gamma N}$, can be  approximated by the 
Caldwell parameterization \cite{Cald} on the basis of experimental data
on photo-production by real photons:
\begin{equation}
\sigma_{\gamma N}(k)=49.2 + 11.1\ln (k)+ \frac{151.8}{\sqrt{k}}
\label{sigmagN}
\end{equation}

\subsection{Analysis of the influence of the muonic events for
different conditions}

In order to estimate the order of magnitude of the different processes
in the case of air showers, we have calculated the mean free path (MFP),
$\lambda$,  of the different effects. Each MFP is inversely
proportional to the corresponding cross section. In fact
$\lambda=m/\sigma$ where $m$ is the mass of the target nuclei.

Figure \ref{fig:mfp}  shows the MFP's  in  \gcmsq \, as a  function of
the kinetic energy of the muon for the cases of MBR, MPP,  emission of
knock-on (KNO) electrons ($\delta$ rays), and muon-nucleus
interaction.  One can compare the MFP's of the different muonic
events with the depth of the atmosphere ($\sim 1000$ \gcmsq \, for
vertical showers and $\sim 9000$ \gcmsq \ for showers with zenith
angle $85^{\circ}$). The influence of such processes in the development of the
shower will be more important for large zenith angles where the total depth
of the shower is bigger and the muonic events have more probability
to take place. 
There is another reason to expect that the influence of the effects
will be more appreciable for large zenith angles: 
The muonic component of the showers at ground level becomes very
important for zenith angles larger than $60^\circ$ (see, for
example, figure 2 in reference \cite{GF}). 

Due to the fact that the MFP's decrease when the initial energy of the
muon is large, it is expected that the influence of the MBR, MPP and
muon-nucleus interactions will be more important under these
conditions.

Figure \ref{fig:mfp} also illustrates that 
the emission of knock-on electrons (KNO) is the most probable process for
energies  smaller than $10^6$ GeV, while  for energies larger than
this value, the MPP dominates. The MFP of this last  process is
about  three orders of magnitude smaller than the ones of MBR and muon
nucleus  interaction.

In the case of MBR, the MFP is  1 or 2 orders of magnitude (for very
inclined  or vertical showers, respectively) greater than the depth of the
atmosphere. Therefore, the total probability of the  process 
during the entire shower path remains always small.

The muon-nucleus interaction competes with the MBR, but,  as it is shown
in figure \ref{fig:kmn} (where the differential cross section of both processes
is plotted versus $v$),  the muon-nucleus interaction has less
probability to  generate hard events and then to produce sub-showers.
Moreover, the average energy loss of the muon-nucleus interaction, that
is,  the integral of  $v$ times the differential cross section
(\ref{sigmagN}), is only about 5\% of the total energy loss.
Due to these facts, the  influence of the muon-nucleus interaction in
the shower development will be less important than the one of MBR.
In consequence, it can safely be assumed that this interaction  will not
appreciably  affect the air shower development and therefore we have
not taken  it into account our  simulations. 

With the aim of analyzing the modifications that MBR and MPP may induce in
the  shower development we have incorporated in AIRES \cite{Aires,Aires2} new
procedures for both processes.
The corresponding algorithms emulate the formulations described in
this section, including all the  details that are relevant for the case of
air showers. The technique used to implement such algorithms employs
in first term very fast approximate calculations of MFP's, using
adequate parameterizations of the corresponding  theoretical
quantities. Then, the exactness of the procedure is ensured by means of
acceptance-rejection tests performed after primary acceptance of the
interactions. As a result, the procedures developed for AIRES do not
increase significantly the computer time required by the general
propagating procedures, while ensuring that the interactions are
treated properly.

\section{Air Shower Simulations}

In order to analyze the influence of MBR and MPP in the development of
air showers initiated by ultra high astroparticles we have performed
simulations using the AIRES program \cite{Aires,Aires2} with different
initial conditions: primary particles (protons, iron nuclei, muons),
primary energies (from $10^{11.5}$ eV to $10^{20.5}$ eV) and zenith
angles (from $0^{\circ}$ to $85^{\circ}$).  We compare results of
simulations where the MBR and MPP have been taken into account with
results obtained in identical conditions but not considering those
muonic processes.  Notice that both, the emission of knock-on
electrons and the muon decay, are {\em always} taken into account.

Unless otherwise specified, the geomagnetic field was not taken into
account in the simulations, in order to avoid large muon deflections
that are present in quasi-horizontal showers.

Hadronic  interactions with primary energy greater than 140 GeV were
processed using the  QGSJET model \cite{QGSJET}, while for energies
below that  threshold, a modified version of the Hillas splitting
algorithm  \cite{Hillas} tuned to match QGSJET predictions at 100 GeV,
was  used. 

Due to the fact that the number of particles in an ultra high  energy
simulation is very large (for example a $10^{20}$ eV shower contains
about $10^{11}$ particles) it is necessary, from the computational
point of view, to  introduce a sampling technique in order to reduce
the number of particles actually simulated.  An extension of the  so
called  thinning algorithm, originally  introduced by Hillas
\cite{Hillas},  is used in AIRES \cite{Aires,Aires2}.  This technique allows
to  propagate all particles  whose energy is larger than  a fixed
energy,  called thinning energy, $E_{thin}$; and   only a small
representative fraction of the total number of  particles is followed
below  this energy. A statistical 
weight is assigned to the accepted particles, which is adjusted to ensure
that the sampling method is unbiased. The thinning algorithm of AIRES
is controlled by two parameters, namely, the thinning energy and the
weight limiting factor, $W_f$. The quality of the sampling improves
when these parameters diminish. The AIRES thinning technique is
explained in detail elsewhere \cite{Aires2}. The thinning energy is
usually expressed in units of the shower primary energy, and in this
case it is named {\em relative thinning}.

\subsection{Evolution of single muons} 

Let us consider first the case of the evolution of a single muon
eventually produced during the development of a given shower. We have 
simulated such muon initiated showers in a representative case: 
Primary energy $10^{14}$ eV and zenith angle $85^{\circ}$. One can
observe that due to MBR and  MPP a muon of such energies may
generate secondary showers. This effect is clearly illustrated
in figure \ref{fig:mld} (a)  where the number of 
electrons and positrons $(e^{+}e^{-})$ is plotted versus the slant depth,
$X_s$.  
Due to the processes of MBR and MPP the number of electrons and positrons
is enlarged with respect to the no MBR-MPP case. 
When these effects are not  taken into account, only KNO and muon
decay can affect the propagation of the muon during all its path.
Notice that the average number of muons is virtually equal to one
during the entire shower development. This can be explained taking
into account that: (1) At  energies of the order of $10^{14}$ eV the muon
decay probability is quite small,  about 1 \% (the mean free path for
decay is approximately $6\times10^{8}$ m, while the length of the
atmosphere  along a $85^{\circ}$  inclined axis is about $10^{6}$ m). (2) The
probability of generating additional muons via decay of pions coming
from photo-nuclear reactions involving secondary gammas is vanishingly small.

In figure \ref{fig:mld} (b) the fraction of energy accumulated by
secondary particles relative to the primary energy, that is
\begin{equation}
f_{E_s} = \frac{E_{prim}- E_{\mu}}{E_{prim}}
\label{fes}
\end{equation}
is plotted versus $X_s$.
When the MBR and MPP effects are not taken into account the muon
almost does not loose energy during all its path ($\sim 0.008$ GeV/(\gcmsq)),
while if such effects  are considered, the muon energy loss rises
up to $\sim 0.3$ GeV/(\gcmsq)  at $10^{14}$ eV. This is  due to the
fact that both MBR
and MPP  have the possibility to produce hard events, responsible for
the more significant losses shown in figure \ref{fig:mld} (b).
On the other hand, when MBR and MPP are disabled, the
muon energy loss comes from the emission of KNO electrons (soft
and hard) and the muon decay that implies a total loss of less than 
0.1 \% of the primary energy, even in the case of horizontal showers.

However, the muon decay may affect the first stage of the average
shower development and, in fact, it is the responsible of the initial
($X_s< 1500$ \gcmsq ) peak of figure \ref{fig:mld} (a) in the
MBR-MPP off case. To understand more clearly the origin of this effect, let
us consider the probability, $P_d$, that the muon of energy $E$ decays before
undergoing any process of knock-on electron emission (For simplicity
we are not taking into account MBR and MPP in this analysis). A
straightforward calculation yields,
\begin{equation} 
P_d= 1 -\int_0^{\infty} \exp\! \left \{- \left[\frac{X}{\lambda_{KNO}} +
\frac{\ell_{X}}{\ell_D}\right] \right\} \;  \frac{dX}{\lambda_{KNO}}
\label{muondecay}
\end{equation}
where  $\lambda_{KNO}$ is the knock-on mean free path in \gcmsq, 
$\ell_D$ is the decay mean free path in meters, $X$ is the matter path
measured from the location of the particle and along its trajectory,
and $\ell_{X}$ is the metric path corresponding to $X$.  $\ell_{X}$
depends also on the location of the muon (represented by its depth $X_s$)
and the atmospheric model used, and $\ell_D$ depends on the muon energy.
Equation (\ref{muondecay}) can be conveniently evaluated numerically
considering a realistic atmospheric model. The results are plotted in
figure  \ref{fig:decayprog} where $P_d$ is represented as a function
of $X_s$.  As expected, $P_d$ is always small and diminishes as long as
$X_s$ grows. At the top of the atmosphere $P_d \cong 6 \times
10^{-4}$.  This means that in a
batch of, say, $10^6$ showers, an average of 600 showers will be initiated by
muon decay. Such showers will be characterized by an initial electron
(or positron) carrying a significant fraction of the primary energy,
and  capable of generating a major electromagnetic shower. These
electromagnetic showers are responsible for the initial peak that
shows up in figure \ref{fig:mld} (a). Notice that the maximum number
of particles for $10^{14}$ eV electromagnetic shower is (roughly)
$10^5$. When averaging, such showers contribute attenuated by a factor
$P_d$. This gives $\langle N_{max} \rangle \cong 6 \times 10^{-4}
\times 10^5 = 60$, result that is in agreement with the
corresponding plot at figure \ref{fig:mld} (a).

\subsection{Evolution of air showers} 
\label{sec:eoas}

The influence of MBR and MPP in the global observables of
air  showers has been exhaustively studied using mainly the
representative case of a proton primary.

Although the relative frequencies of MBR and MPP in all cases are small
compared with  other muonic events like  KNO, in some  conditions
the influence of these processes in the development of the
shower is not negligible (For example, as we have just mentioned in the last
paragraph, the MBR and MPP may generate sub-showers).
Figure \ref{fig:rfme} displays the percentage  of muonic  events for
$10^{19}$ eV proton showers with zenith angles of  $45^\circ$ (solid
lines)  and  $85^\circ$ (dashed lines). The bars
correspond, respectively,  to KNO, MBR, MPP,  and muon decay (MDY).
As it is expected,  the KNO processes always account for the largest
frequency, 96.36 \% (96.64 \%) in the $45^\circ$ ($85^\circ$) case.
The other processes are by far less frequent:
2.29 \%,  1.32 \% and 0.03 \% for MDY, MPP and MBR respectively
($45^\circ$ case). Comparing the percentages of muonic events in the 
$85^\circ$  case against the $45^\circ$ ones, it can be seen that both
MPP  and MBR rates  are slightly  increased  (about  3 \% and 0.05 \%
respectively),  while the MDY relative frequency diminishes. 

In figure \ref{fig:erme} the energy distributions of the different
muonic  events are represented. The initial conditions of the
shower are the same as in figure \ref{fig:rfme}. In agreement with the
MFP's of figure \ref{fig:mfp}, MBR and MPP occur, in average, at
relative large energies. On the other hand, and as expected, MDY takes
place at lower energies. Figure \ref{fig:erme} shows that the energy
spectrum  of the muons that undergo the studied events moves
slightly  towards large energies if the zenith angle of the shower is
increased.

We have also studied the frequency  of the muonic events as a function
of the primary energy. The results are  shown in figure
\ref{fig:percentmuevents} where the percentages of muonic events are
plotted versus the primary energy. The main characteristic of these
plots are the following:
For primary energies above $10^8$ GeV, all the percentages remain
practically invariant.
The KNO effect is always the one with maximum relative percentage.
The MDY presents a noticeable dependence with the primary energy in
the region below $10^8$ GeV (from 10 \% at $10^{3}$ GeV down to 0.4 \%
at $10^8$ GeV). 
On the other hand, the MPP grows with the primary energy although the
difference between extremes is less significant than in the MDY case
(from  0.9 \% at  $10^{3}$ GeV  up to 3.5 \% at  $10^{8}$ GeV).
MBR behaves similarly than MPP, but this fraction is about two orders
of magnitude smaller than the MPP one.

The particular behavior of these fractions can be explained
considering the characteristics of the energy distribution of the
different muonic events, plotted in figure \ref{fig:distegykpbdqgs}
for several primary energies. All the spectra can, in principle,
extend up to the primary energy of the shower. When the primary
energy is less than $\sim 10^{14}$ eV, this cutoff is clearly
visible in the plots of figure \ref{fig:distegykpbdqgs}. In these
cases, the muons generated during the shower have a non negligible
decay probability. For primary energies above $10^{14}$ eV, the energy
distribution of muons broadens, but the spectrum of decaying
muons remains bounded in the region $E\lesssim 10^{12}$ eV, due to the
fact that the
decay probabilities become very small for energies above that
limit. As a consequence, the total fraction of decaying muons
diminishes progressively with the primary energy, as shown in figure 
\ref{fig:percentmuevents}. 

When the primary energy is much larger than $10^{14}$ eV, the energy
distribution of muons is concentrated in the region of energies lower
than that indicative value, and only a small tail extends to higher
energies. As a consequence, most significant part of the energy
distributions for all the muonic events become almost independent of
the primary energy, and so the fractions plotted in figure
\ref{fig:percentmuevents} do not present important changes at the
highest primary energies.

The MPP relative fraction depends mainly on the number of high
energy muons, which rises significantly with the primary energies
for $E_{prim}\lesssim 10^{15}$ eV and stabilizes above that energy. 

The very small variations in the fractions of figure
\ref{fig:percentmuevents} at the highest energies (increase of MDY and
decrease of MPP fractions) can be regarded as secondary effects of the
variations of the characteristics of the hadronic processes that take
place at the beginning of the shower development. Cross sections and
multiplicity of hadronic collisions rising with energy are some of the
aspects that need to be taken into account in this sense.  A detailed
discussion on the characteristics of the inelastic hadronic
collisions is beyond the scope of this work; the interested reader
can consult reference \cite{Anchordoqui}.

\subsubsection{Longitudinal development}

The following paragraphs contain a description of the modifications
induced on different shower observables due to the MBR and MPP
effects. We consider first the case of $3\times 10^{20}$ eV proton
showers inclined $85^\circ$.

One of the most evident modifications induced by the MBR and MPP effects
is the increase of the size of the residual electromagnetic  shower
produced during the late stages of the shower development (well beyond
the shower maximum).

This shows up clearly in  figure
\ref{fig:p6_z85_3e20_elctrongammaerrnew}  where the
number of gammas (a1) and electrons and positrons (b1) are plotted
against $X_s$. Notice that, accordingly to our calculations, there are no
visible  differences in the position of the maximum of the shower
($X_{max}$).
In order to show  the increase of the electromagnetic shower it is
convenient to define  the relative difference between the cases where
the MBR and MPP are or  are not taken into account, that is: 
\begin{equation}
\Delta = \frac{\rm{N_{MBR/MPP \;
On}}}{\rm{N_{MBR/MPP \; Off}}} -1.
\label{Delta}
\end{equation}
$\Delta N_{\gamma}$ and $\Delta N_{e}$ have been plotted in
figure  \ref{fig:p6_z85_3e20_elctrongammaerrnew} (a2) and (b2),
respectively. For clarity, these plots include only the tail of the
showers ($X_s > 2200$ \gcmsq). It can be noticed that the
relative increase of the number of gammas and electrons is about 20 \% at
the very late stages of the shower development.

Similar plots describing the development of the average energy of
gammas and electrons and positrons are displayed in figure
\ref{fig:p6_z85_3e20_elctrongammaerrenergnew}. The fact
that the relative increments of the energies are similar to the
corresponding relative increments of particle numbers indicates that
the energies of the electromagnetic particles are not
substantially modified by the inclusion of the MBR and MPP effects, as
expected.

The influence of  MBR and MPP is less significant on the
muonic component: The number of muons during the development of the
shower practically does not change if these effects are considered.  
The longitudinal development of muon energy  appears in figure
\ref{fig:pldm} (a). This plot shows that the energy of the muons
diminishes  about  3 \% at the tail of the shower if the MBR and MPP
effect are enabled. 
It is also observed  that the sum of the energies of all muons divided
by the  average number of muons,
\begin{equation}
\xi_{\mu}= \frac{E_{\mu}}{N_{\mu}}
\label{ximu}
\end{equation}
is not significantly modified  when the effects are considered for the
primary energy mentioned above (see figure \ref{fig:pldm} (b)).

We have also investigated whether or not  the MBR and MPP generates
modifications in the shower front arrival time profile for  different
particles of the shower (muons, electrons and gammas). We have not found
any  significant alteration when comparing the cases when  the effects
of  MBR and MPP are or not taken into account.

We have studied the modifications introduced by the MBR and MPP
effects for different primary energies. The influence of these
processes in the electromagnetic component of the showers with smaller
primary energy  is similar to the case of $3\times 10^{20}$ eV,
described above.

In figure \ref{fig:p6_z85vs_snewmuon}, $\xi_{\mu}$ is plotted versus
$s=X_s/X_{max}$ for different primary energies. All the curves show a
similar behavior: (i) In the region $0 <s <1$ $\xi_{\mu}$ decreases
with $s$, direct consequence of the multiplicative processes that take
place in this phase and increase the number of shower secondaries,
thus reducing the average energy per secondary. (ii) For $s>1$, low
energy muons decay progressively, and therefore the mean muon energy
is shifted as long as $s$ grows. When comparing the curves
corresponding to different primary energies, it is possible to see
that $\xi_{\mu}$ increases monotonically as long as the primary energy
decreases from $E_{prim}= 3 \times 10^{20}$ eV (a) to $E_{prim}\cong
10^{14}$ eV (c); and decreases when the energy continues decreasing
below $10^{14}$ eV (curves (c) and (d) for $3 \times 10^{11}$ eV; no
intermediate cases were plotted for simplicity). These behavior is
consistent with the characteristics of the energy distributions of
figure \ref{fig:distegykpbdqgs}, already explained at the beginning of
section \ref{sec:eoas}: The low energy range (curves (c) and (d)) are
characterized by muon spectra bounded by the primary energy and thus
significantly changing when it varies, and with a mean value
increasing with the primary energy. On the other hand, the high energy
range (curves (a) to (c)) is characterized by muon spectra weakly
correlated with the primary energy, and enhanced fraction of low energy
muons at the highest primary energies.

When comparing the curves corresponding to the cases MBR-MPP disabled
(dashed lines) and enabled (solid lines), it is possible to notice that
the differences between pairs of curves is always small, with a maximum of 5
\%  for curve (c) at $s=10$. In general $\xi_{\mu}$ decreases when MPP
and MBR are switched on. However, a critical combination of event
probabilities (see figure \ref{fig:percentmuevents}) determines that
$\xi_{\mu}$ remains unchanged or is slightly increased for primary energies
around $10^{15}-10^{17}$ eV.

From figure \ref{fig:p6_z85vs_snewmuon} one can see that in the region
around the shower maximum, that is, where $s$ ranges approximately
between 0 and 2, $\xi_{\mu}$ practically does not present significant
changes when comparing the cases where the MBR and MPP are enabled or
disabled.  On the other hand, progressively significant differences
can appear for $s$ larger than 2.  $s$ can be regarded as a measure of
the stage of the shower development, ranging from 0 at the top of the
atmosphere to a final value $s_g$ at ground which depends on the
zenith angle.  $s_g$ is a measure of the quantity of matter the shower
has to pass through from its beginning until reaching the ground
level.  From the plot in figure \ref{fig:p6_z85vs_snewmuon}, it can be
inferred that when $s_g$ is less than 2, that is, for zenith angles
less than $\sim 45^\circ$ in the conditions of our simulations, there
will be no noticeable modifications on the shower development due to
MBR and MPP.

We have also analyzed the influence of MBR and MPP in the showers
initiated by different primary particles like, for example, iron nuclei. 
We have observed  that the modifications that  MBR and MPP
introduce in the late stages of the shower development have
approximately the same characteristics of the ones introduced for a proton
shower for the case of the electromagnetic component of the mentioned
shower. For the case of muons observables the
differences are less significant. For example, $\Delta_{E_{\mu}}$
reaches  a maximum of 1.5 \% in the late stages of the shower
development for the case of iron shower of $3\times 10^{20}$ eV and
$85^{\circ}$ of zenith angle while for proton showers, in the same
initial conditions, $\Delta_{E_{\mu}}$ is  3 \%.

\subsubsection{Lateral distributions} 

We have also studied the  distributions of the particles at ground
level (lateral distributions).
In the case of very inclined showers, the intersection with the ground
plane occurs well beyond the shower maximum, and lateral distributions
are somewhat different with respect to the ``typical'' distributions
corresponding to showers with small zenith angles. Among other
differences, we can mention: (i) Substantially smaller number of
particles. (ii) The densities of the electromagnetic particles are of
the same order of magnitude than  the density of muons (In the
case of quasi vertical  showers, the muons account for only about 1
\% of the ground particles).

In figure \ref{fig:lateral} the densities 
of $\gamma$, $e^+e^-$, and $\mu^{+}\mu^{-}$, are plotted versus the
distance to the shower axis, for the case of $10^{19}$ eV proton
showers inclined $70^{\circ}$. The ground level is located at 875
\gcmsq \ . The analysis of  the data shows that the number of $\gamma$ and 
$e^+e^-$ is slightly modified $-$when MBR and MPP are taken into
account$-$ near the shower axis, while the lateral distribution of
muons remains virtually unaltered when the MBR and MPP interactions
are  enabled. It is worthwhile to mention that in this case the
geomagnetic field is taken into account in order to simulate a real
situation. In fact, we have chosen the conditions corresponding to the
site of El Nihuil (Mendoza, Argentina) with the aim of studying the
characteristics of showers to be measured by the future Auger
Observatory \cite{Augerwww} that is currently being constructed at that
site.

The measurement of the lateral distributions of particles at ground
generated by ultra high energy cosmic rays is usually performed by
means of water  \u{C}erenkov detectors. Such devices are sensible to
both  electromagnetic particles and muons, and the signal they produce is
the sum of both components. 
The signal produced by any particle hitting a water  \u{C}erenkov
detector must be estimated by a specific Monte Carlo simulation which
takes into account the characteristics of the detectors. The detailed
simulation of water  \u{C}erenkov detectors is beyond the scope of our
work; we have instead evaluated estimations of these signals using a
direct conversion procedure that retrieves {\em average} signals
\cite{Kutter}.  Such averages were evaluated on the basis of
simulations  performed with the AGASIM program \cite{Agasim}.

We have plotted in figure \ref{fig:signal} the  ratio between
the  electromagnetic and the muonic component of the detector signal, as
a function of the distance to the shower axis, for the cases when MBR
and MPP are or not taken into account in the simulation of the showers.

The increase of the size of the residual electromagnetic shower that
takes place when MBR and MPP are enabled, produces a larger signal
close to the shower axis, as it clearly shows up from the plots of
figure \ref{fig:signal}. For distances to the axis less than 30 m the
electromagnetic to muon signal ratio increases slightly when the MBR
and MPP  are switched on. On the other hand, this
increment is smaller for larger distances, and is virtually negligible
beyond 200 m from the shower axis.

It is worthwhile to remark that the fact that the main modifications in
the electromagnetic to muon signal ratios are concentrated in a narrow
zone around the shower axis makes  it unlikely that the incorporation
of MBR and MPP in the air shower simulation engine will significantly
affect the results obtained in references  \cite{Zas,Zas2} where the
measurements of inclined showers performed at the Haverah Park
experiment \cite{HavPark} are analyzed with the help of air shower
simulations using AIRES without taking into account MBR and MPP. 

Notice also that the data plotted in figure \ref{fig:signal}
corresponding to the case when the MBR and MPP are switched off can be
compared with the corresponding data presented in reference
\cite{Zas}. It is easy to see that there is a good qualitative
agreement between the two sets of data and that the small differences
between the two works most  probably come from differences in the
ground altitude and/or
specific  parameters used to calculate the detector responses.
 
\section{Conclusions}

We have studied the influence of the MBR and MPP in the development of
air showers initiated by ultra high energy astroparticles. We have 
incorporated in the AIRES air shower simulation system the corresponding
procedures to emulate these effects and  have then performed
simulations in a  number of  different initial conditions.

The analysis of the evolution of a single muon indicates that such
particle can eventually generate  secondary showers after undergoing 
hard  MBR and MPP processes. This indicates clearly that these
interactions  cannot be simulated accurately as continuous energy loss
processes.  

For $3 \times 10^{20}$ eV proton and iron primaries the main
modifications  introduced by MBR
and MPP  affect the  electromagnetic component of the showers.
The number and energy of gammas and electrons increase significantly
in the  late stages of the shower development  (well beyond the shower
maximum), but the mentioned effects do not  generate visible changes
in the position of $X_{max}$.

The changes generated by MBR and MPP for muon observables are less
significant: The number of muons practically
does not change and their energies diminish about 3 \% (1.5 \%) for
the case of proton (iron) showers  at the tail of the shower.

The shower front arrival time profile does not present modifications
due to the MBR and MPP processes.

For primary energies below $3\times 10^{20}$ eV the modifications in
the electromagnetic shower induced by MBR and MPP are qualitatively
similar to the ones described in the preceding paragraphs. In the case
of muon observables like $\xi_{\mu}$ we have found, in the entire
range of primary energies that was considered, small variations
due to MBR and MPP. Such small modifications correspond, in general,
to decrements in the average muon energies when MBR and MPP are
switched on. However, critical combinations of
event probabilities determine that  $\xi_{\mu}$ can remain unchanged
or be slightly increased for primary energies around $10^{15}-10^{17}$ eV.

The fact that the alterations in the electromagnetic showers are only
significant in the late stages of the development of the showers, i.e.,
$X_s>X_{max}$, implies that in normal conditions there will be no
visible changes in the electromagnetic shower at ground level, for
showers with zenith angle less than $45^{\circ}$.

For showers with zenith angles larger than $60^{\circ}$, the MBR and
MPP processes are responsible for an increment of the density of
electromagnetic particles at ground, which is most important in a
narrow region around the shower axis. In connection with this result,
we have also found that the signal produced by  \u{C}erenkov detectors
will be also larger near the shower axis if the mentioned effects are
taken into account.

\section{Acknowledgments}

This work  was partially supported by Consejo Nacional de
Investigaciones Cient\'ificas y T\'ecnicas,  Agencia Nacional de
Programaci\'on Cient\'ifica of Argentina, FOMEC program, and Fundaci\'on
Antorchas. We wish to thank Professors H. Fanchiotti and
C. A. Garc\'ia Canal (University of La Plata) for enlightening discussions. 
We are also indebted to Dr. C. Hojvat (Fermilab, USA) for helping us
to access many of the  references cited  in this work.

\def\journal#1#2#3#4{{\em #1,\/} {\bf #2}, #3 (#4)}

\clearpage
\def\epsfig#1{\epsfbox{#1}}
\begin{figure}
\begin{displaymath}
\hbox{\epsfig{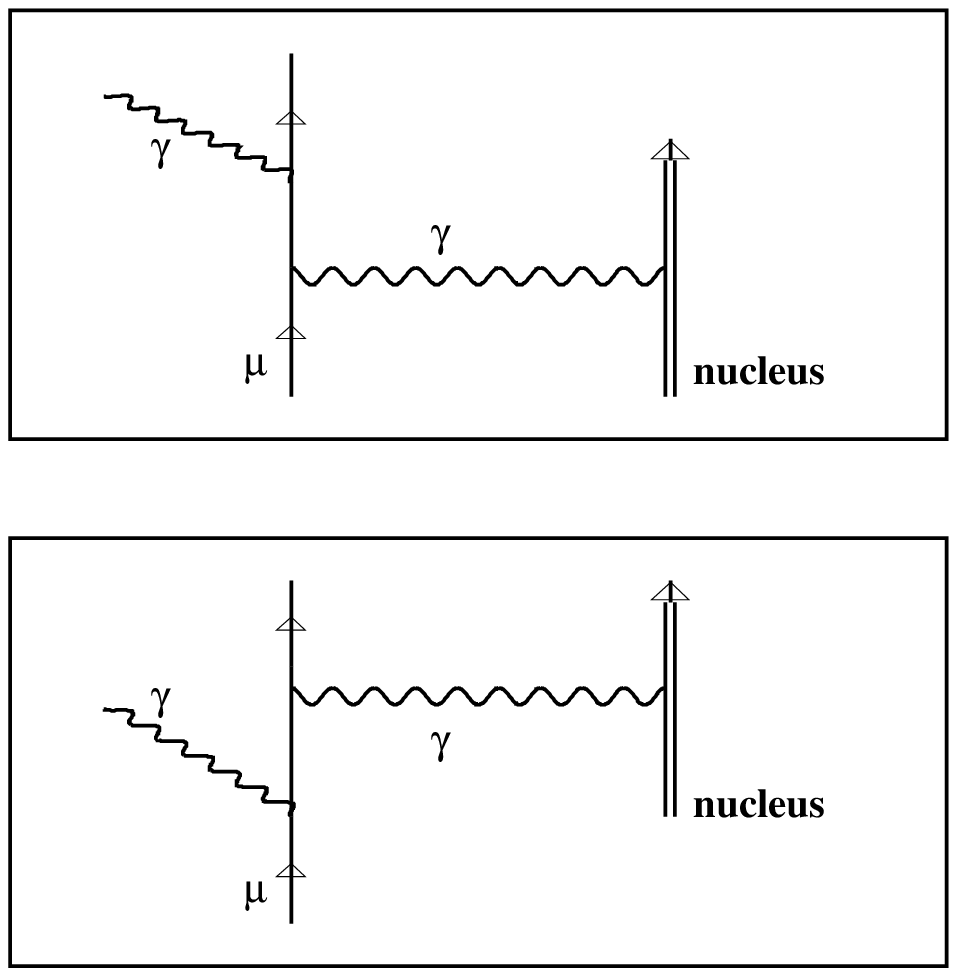}}
\end{displaymath}
\caption{Feynman diagrams for MBR corresponding to the lowest
significant order of perturbation theory.}
\label{fig:fdmb}
\end{figure}
\clearpage
\begin{figure}
\begin{displaymath}
\hbox{\epsfig{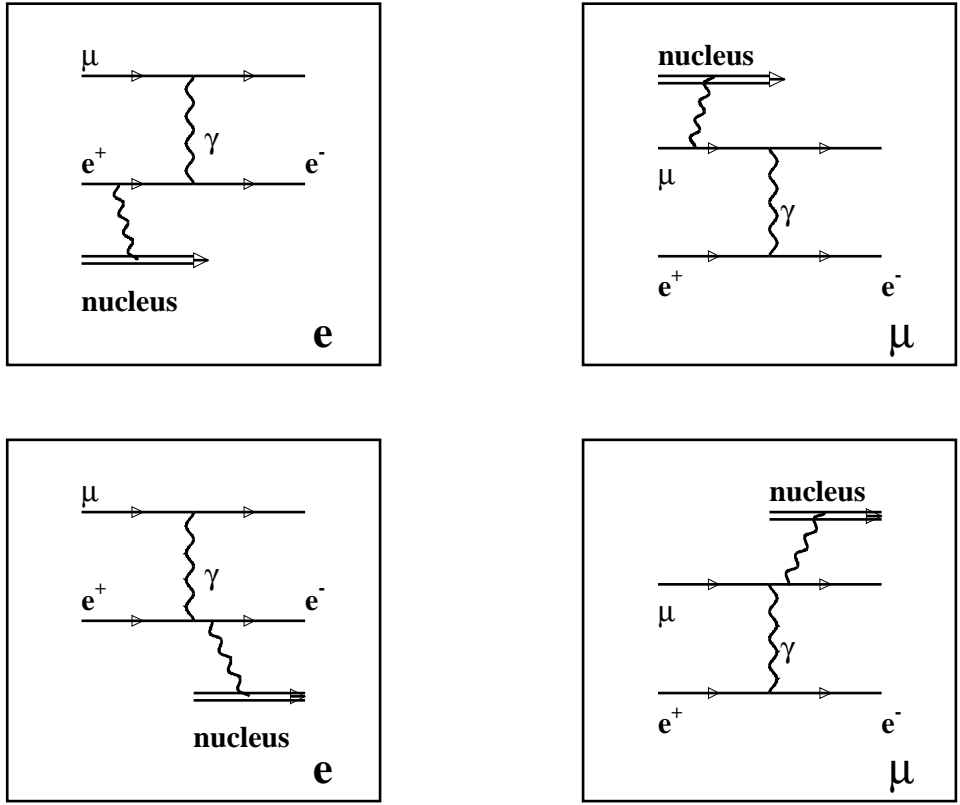}}
\end{displaymath}
\caption{Same as figure 1, but for MPP. In this case the relevant
lowest order of perturbation theory is the 4th.}
\label{fig:fdmpp}
\end{figure}
\clearpage
\begin{figure}
\begin{displaymath}
\epsfig{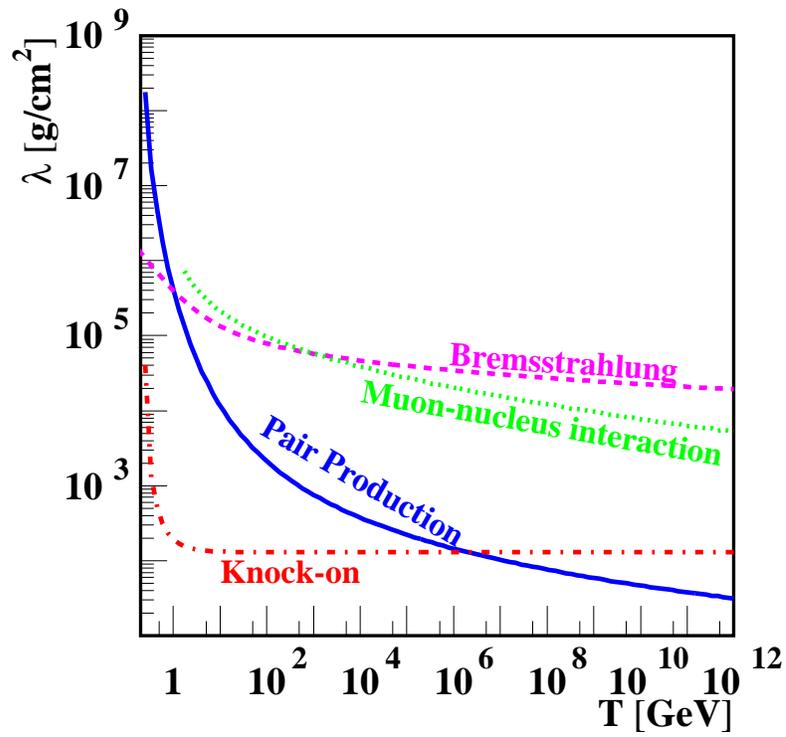}
\end{displaymath}
\caption{Mean free path in air for the different muonic interactions,
plotted versus the initial kinetic energy of the muon. In all cases
the cut off energy is $E_c= 10$ MeV.}
\label{fig:mfp}
\end{figure}
\clearpage
\begin{figure}
\begin{displaymath}
\hbox{\epsfig{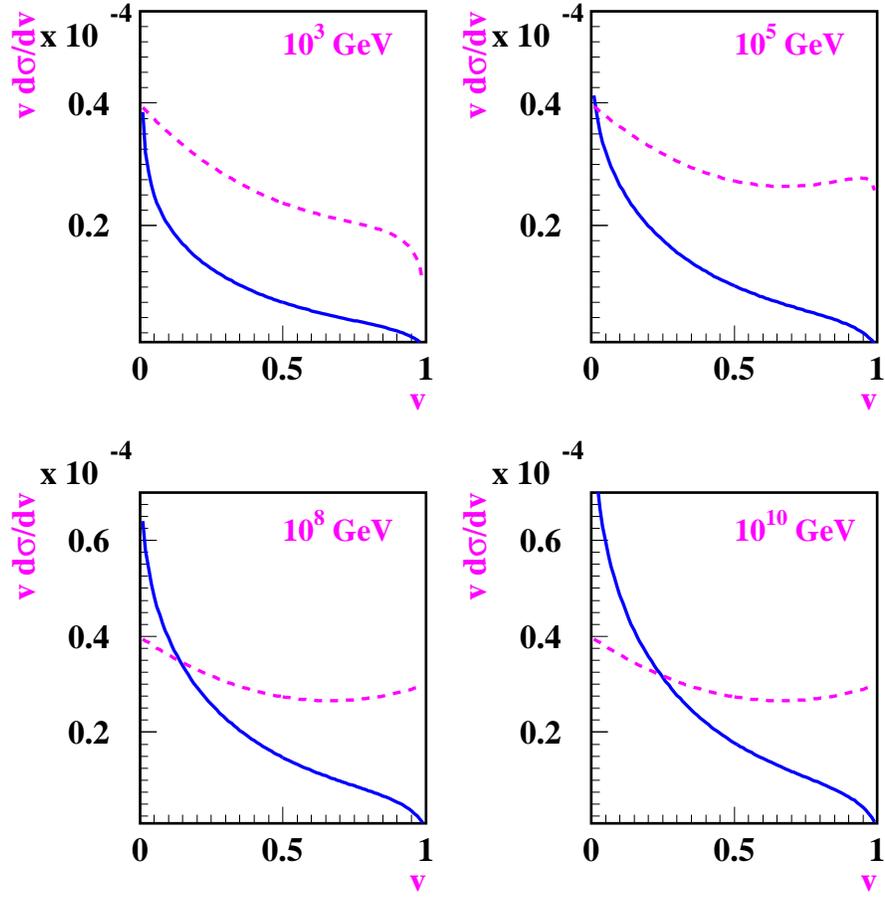}}
\end{displaymath}
\caption{ Differential cross section of MBR (dashed lines) and muon-nucleus
interaction (solid lines) versus $v$ for different muon energies.}
\label{fig:kmn}
\end{figure}
\clearpage
\begin{figure}
\begin{displaymath}
\hbox{\epsfig{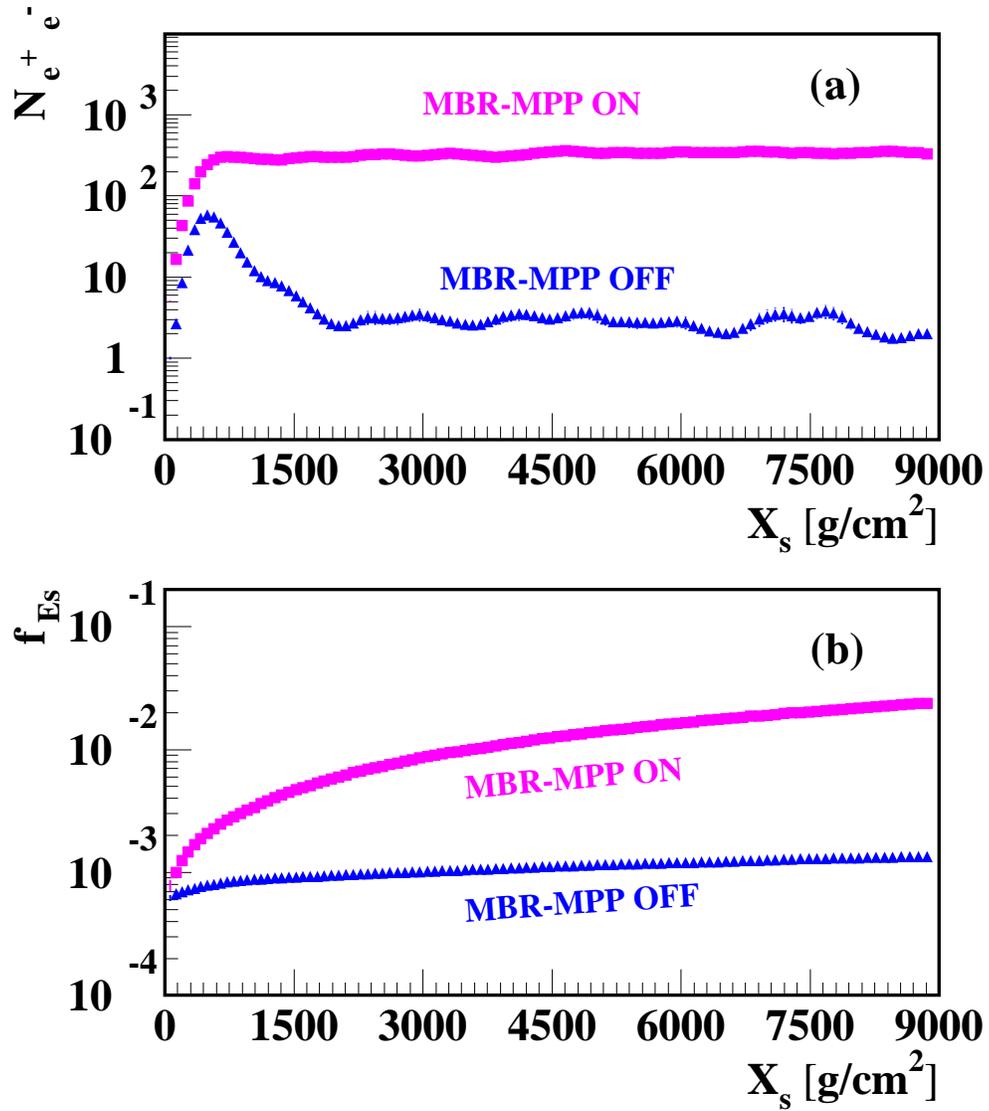}}
\end{displaymath}
\caption{Longitudinal development of showers initiated by
$10^{14}$ eV  muons ($85^{\circ}$ zenith angle) plotted versus $X_s$.
(a) Average number of electrons and positrons. 
(b) Fraction of energy accumulated by secondary particles relative to
the primary energy ($10^{14}$ eV). The averages were performed using a
sample of $10^5$ ($5\times 10^5$) showers in the ON (OFF) case,
simulated  with $10^{-6}$ relative  thinning and
weight limiting factor 3. The geomagnetic field is not taken into
account in these simulations.}
\label{fig:mld}
\end{figure}
\clearpage
\begin{figure}
\begin{displaymath}
\hbox{\epsfig{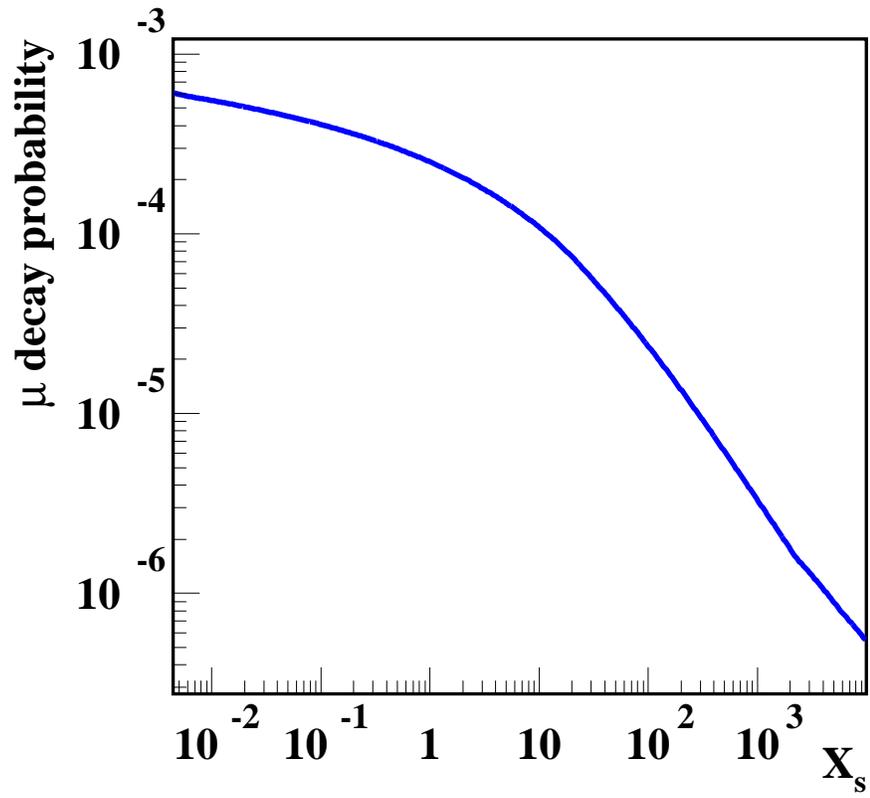}}
\end{displaymath}
\caption{Probability of muon decay in the first interaction versus
$X_s$ (equation (\ref{muondecay})).
The muon primary energy is $10^{14}$ eV.}
\label{fig:decayprog}
\end{figure}
\clearpage
\begin{figure}
\begin{displaymath}
\hbox{\epsfig{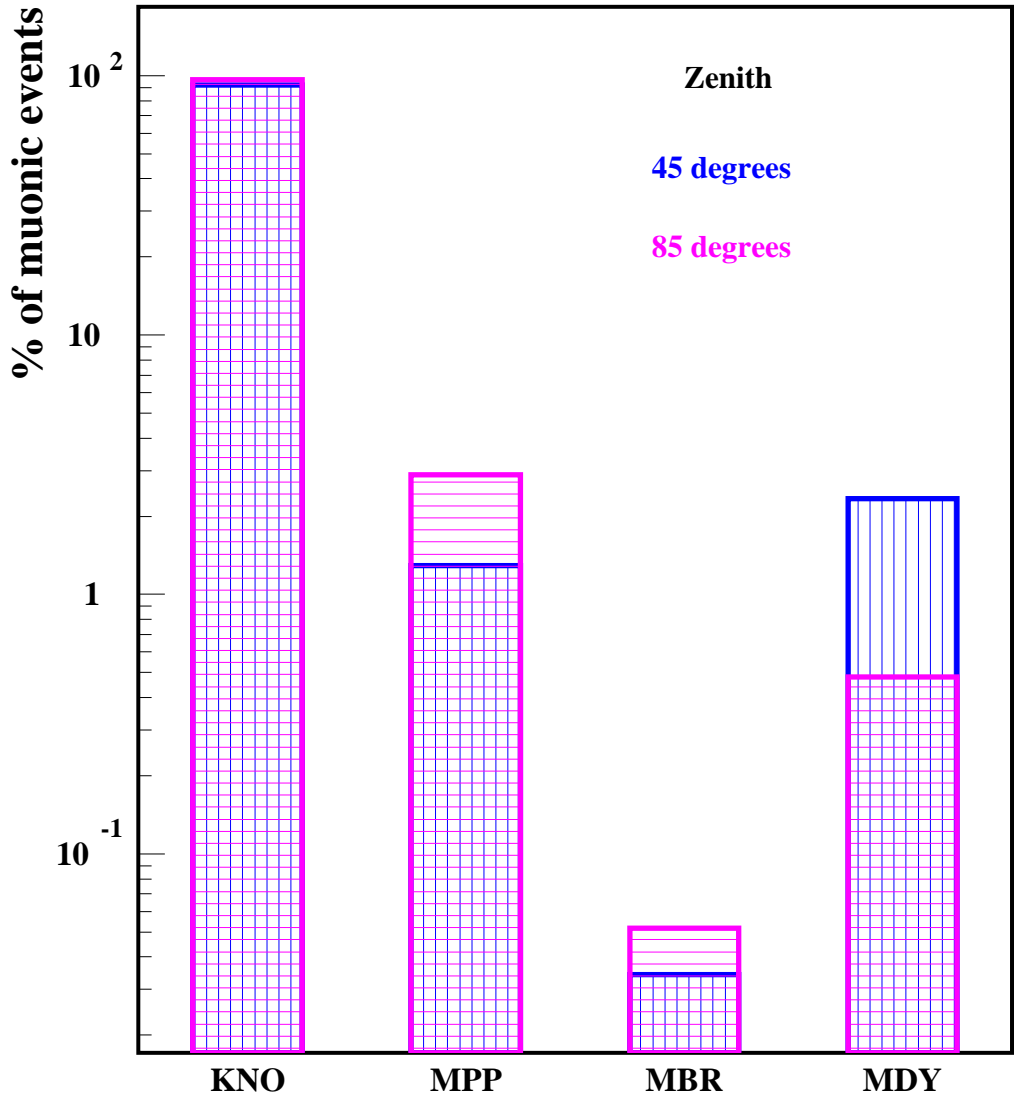}}
\end{displaymath}
\caption{
Percentage of muonic events for $10^{19}$ eV proton
showers (see text). These data represent an average coming
from samples of 50 showers simulated with $10^{-5}$ relative thinning
and weight factor 5.}
\label{fig:rfme}
\end{figure}
\clearpage
\begin{figure}
\begin{displaymath}
\hbox{\epsfig{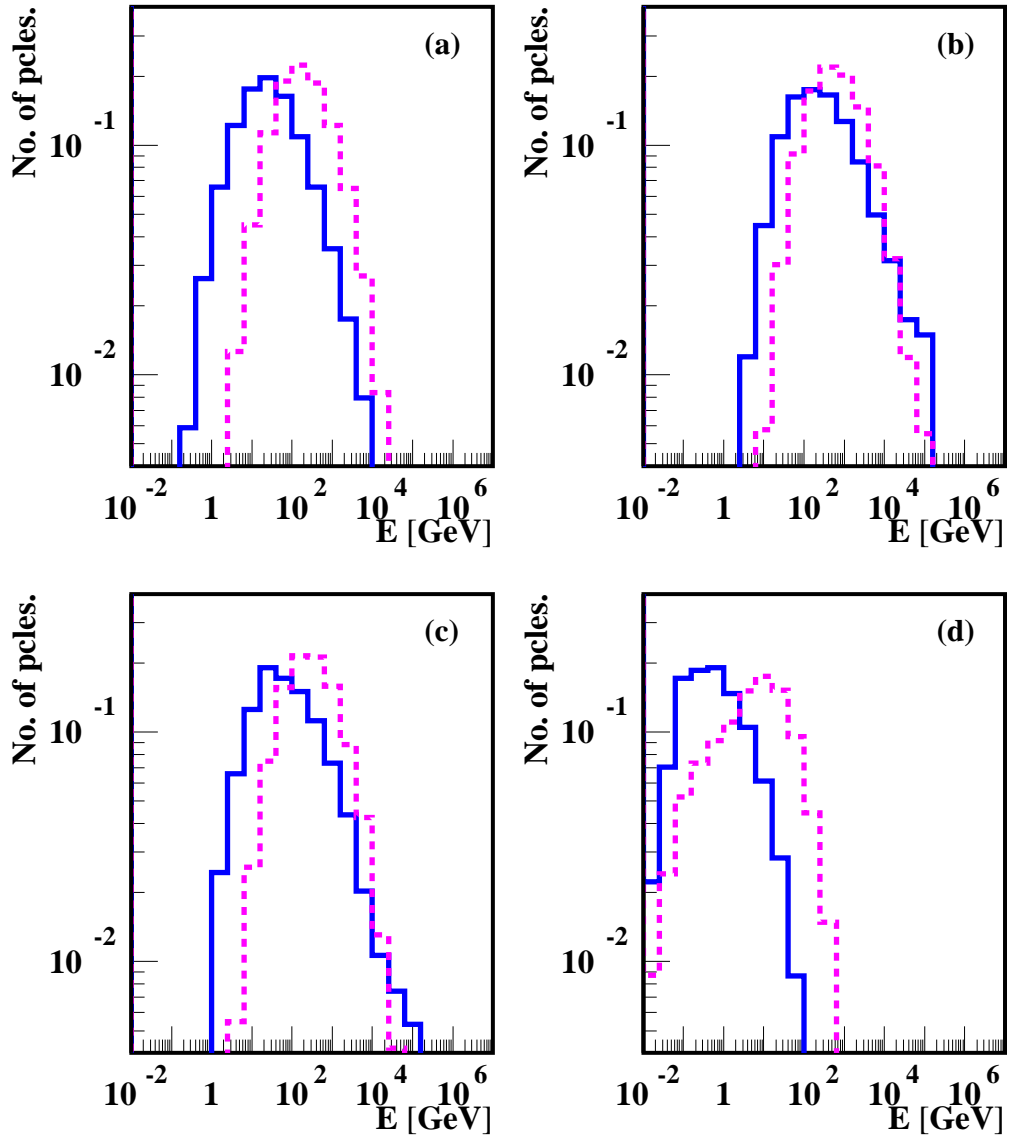}}
\end{displaymath}
\caption{Energy distribution of  muonic events in the same
conditions as in figure \ref{fig:rfme}. Solid (dashed) lines
correspond to $45^{\circ}$ ($85^{\circ}$) zenith angle. (a) is for KNO
processes, (b) MPP, (c) MBR,  and (d) MDY.}
\label{fig:erme}
\end{figure}
\clearpage
\begin{figure}
\begin{displaymath}
\hbox{\epsfig{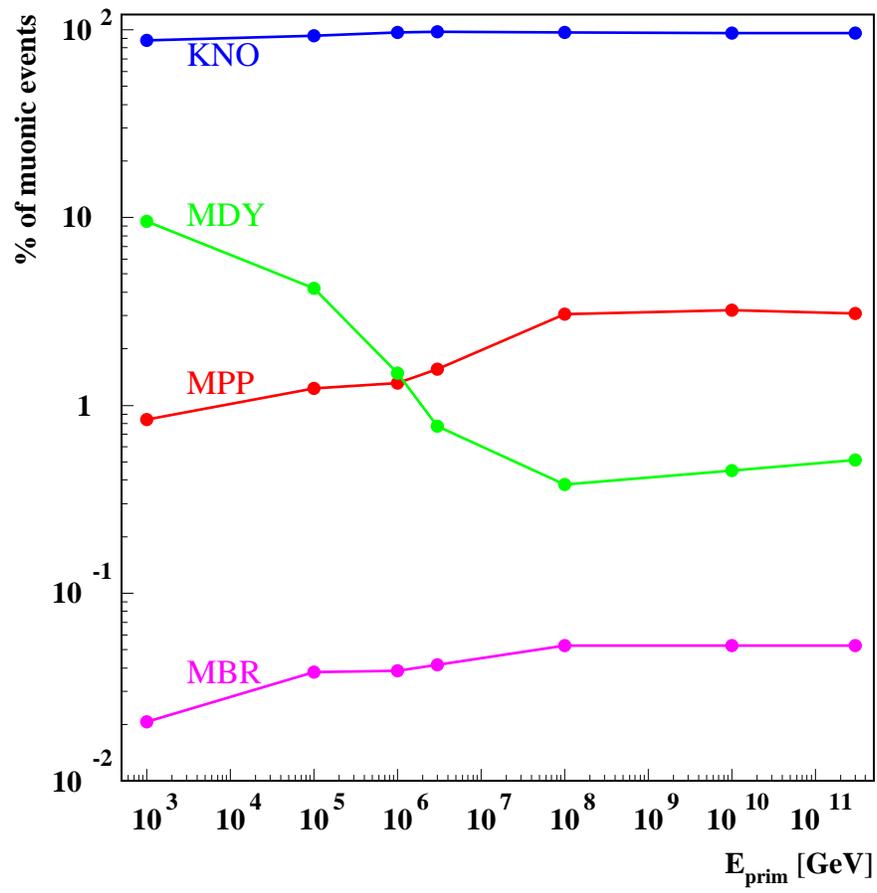}}
\end{displaymath}
\caption{Percentage of muonic events versus the  energy of the
primary particle, for proton showers inclined 85 degrees, simulated in
the same conditions as in figure \ref{fig:rfme}.}
\label{fig:percentmuevents}
\end{figure}
\clearpage
\begin{figure}
\begin{displaymath}
\hbox{\epsfig{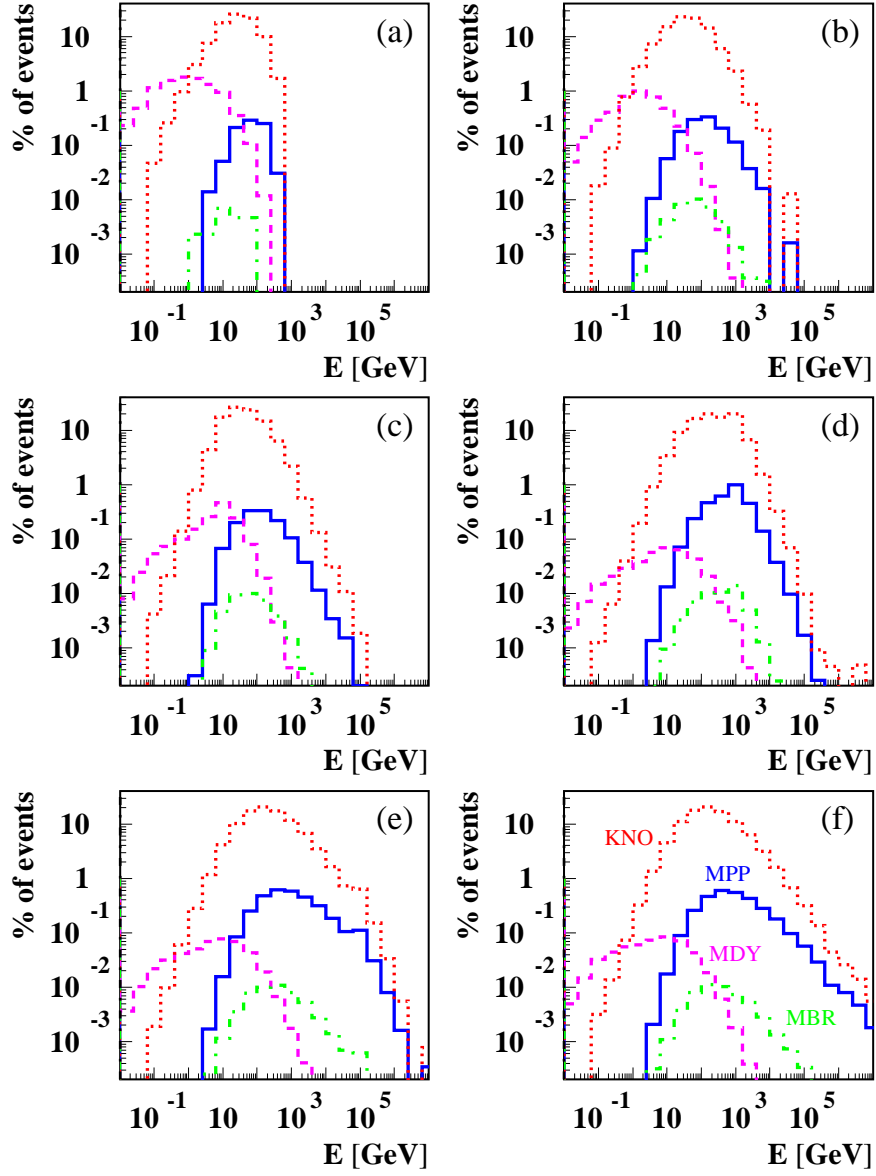}}
\end{displaymath}
\caption{Energy distribution  of muonic events for different
primary energies: (a) $10^{12}$ eV, (b) $10^{14}$ eV, (c)
$10^{15}$ eV, (d) $10^{17}$ eV, (e) $10^{19}$ eV, 
and (f) $3\times 10^{20}$ eV. For clarity, curve labels are indicated
only in (f), and the same pattern applies to all the graphs: Dotted lines
correspond to KNO; solid lines to MPP; dashed lines to MDY; and
dotted-dashed lines to MBR.
The conditions of the simulations are the same as in figure
\ref{fig:percentmuevents}.
}
\label{fig:distegykpbdqgs}
\end{figure}
\clearpage
\begin{figure}
\begin{displaymath}
\hbox{\epsfig{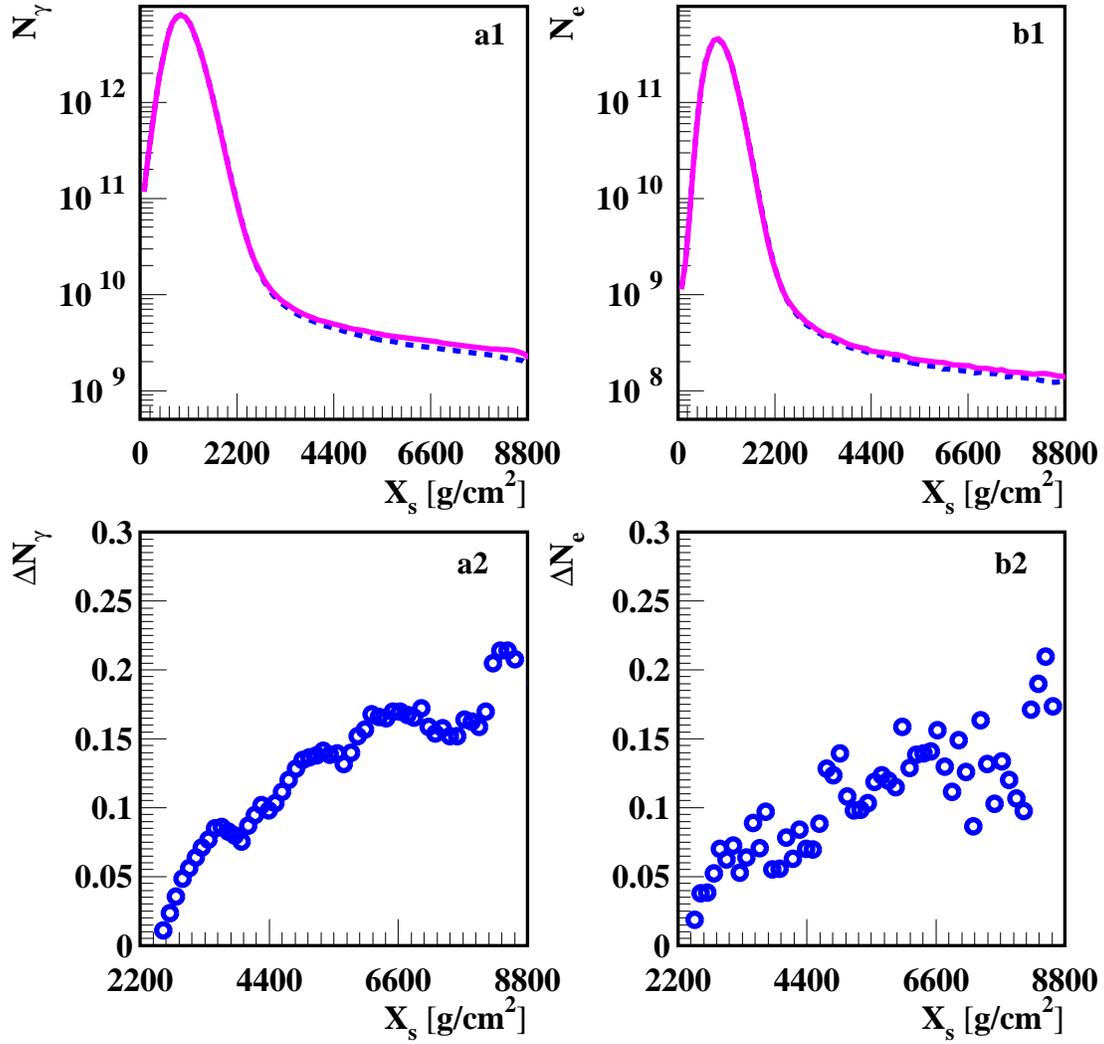}}
\end{displaymath}
\caption{Average longitudinal development of gammas (a1) and the relative
difference $\Delta N_{\gamma}$ (a2) (see text), and average longitudinal
development of electrons and positrons (b1) and the relative
difference $\Delta N_{e}$ (b2)  versus $X_s$, for $3\times
10^{20}$ eV proton showers with a zenith angle of $85^{\circ}$. The
solid  (dashed) lines corresponds to the case where the MBR and MPP
are (are  not) taken into account. The averages were performed using, at
each case, 200  showers simulated with $10^{-6}$ relative thinning and
weight factor 3. The geomagnetic field is not taken into account
during the simulations.}
\label{fig:p6_z85_3e20_elctrongammaerrnew}
\end{figure}
\clearpage
\begin{figure}
\begin{displaymath}
\hbox{\epsfig{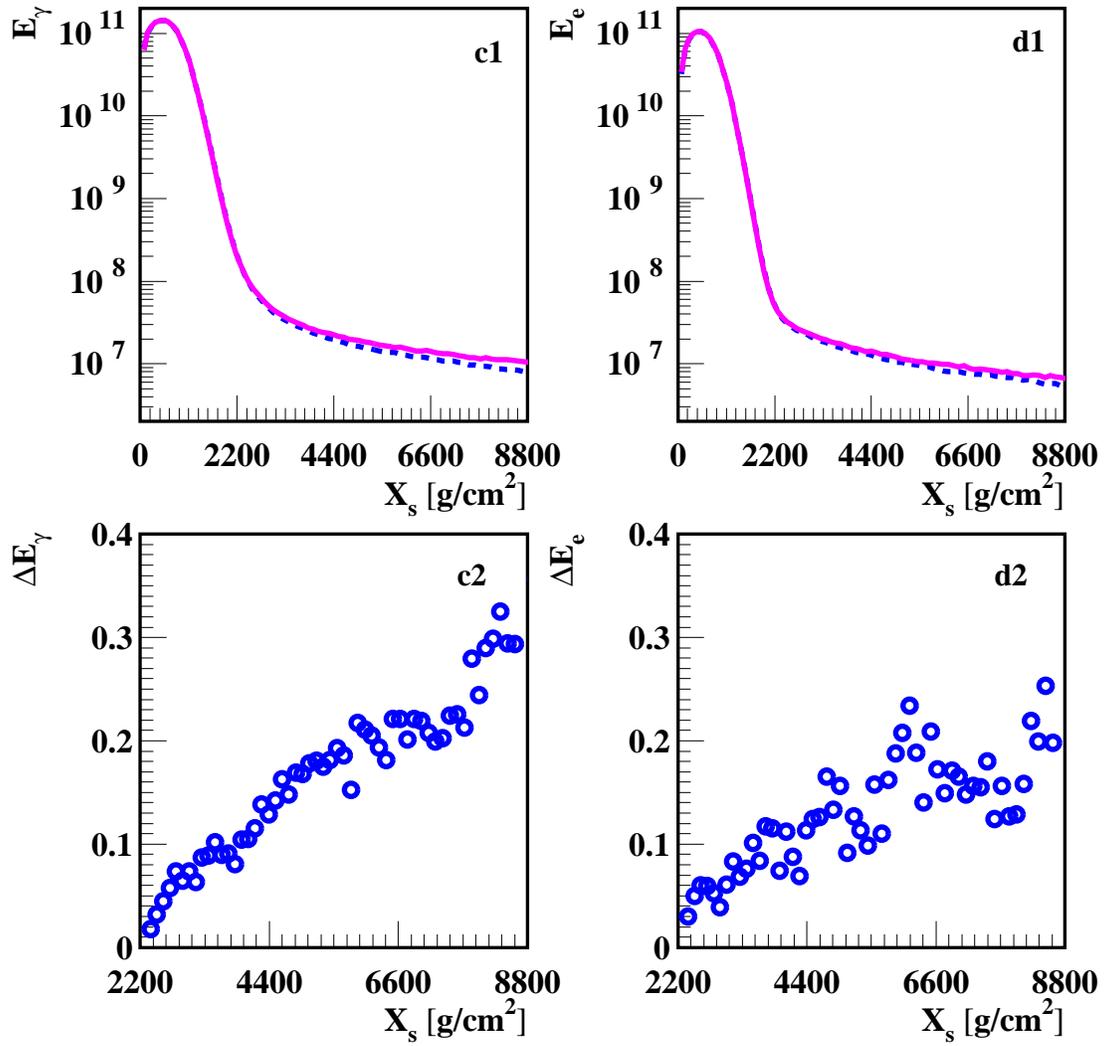}}
\end{displaymath}
\caption{Average longitudinal development of the energy of gammas (c1) and
the  relative difference $\Delta E_{\gamma}$ (c2), and average longitudinal
development of the energy of electrons and positrons (d1) and the relative
difference $\Delta E _e$ (d2)  versus $X_s$. The initial conditions
are the same as in figure \ref{fig:p6_z85_3e20_elctrongammaerrnew}.
}
\label{fig:p6_z85_3e20_elctrongammaerrenergnew}
\end{figure}
\clearpage
\begin{figure}
\begin{displaymath}
\hbox{\epsfig{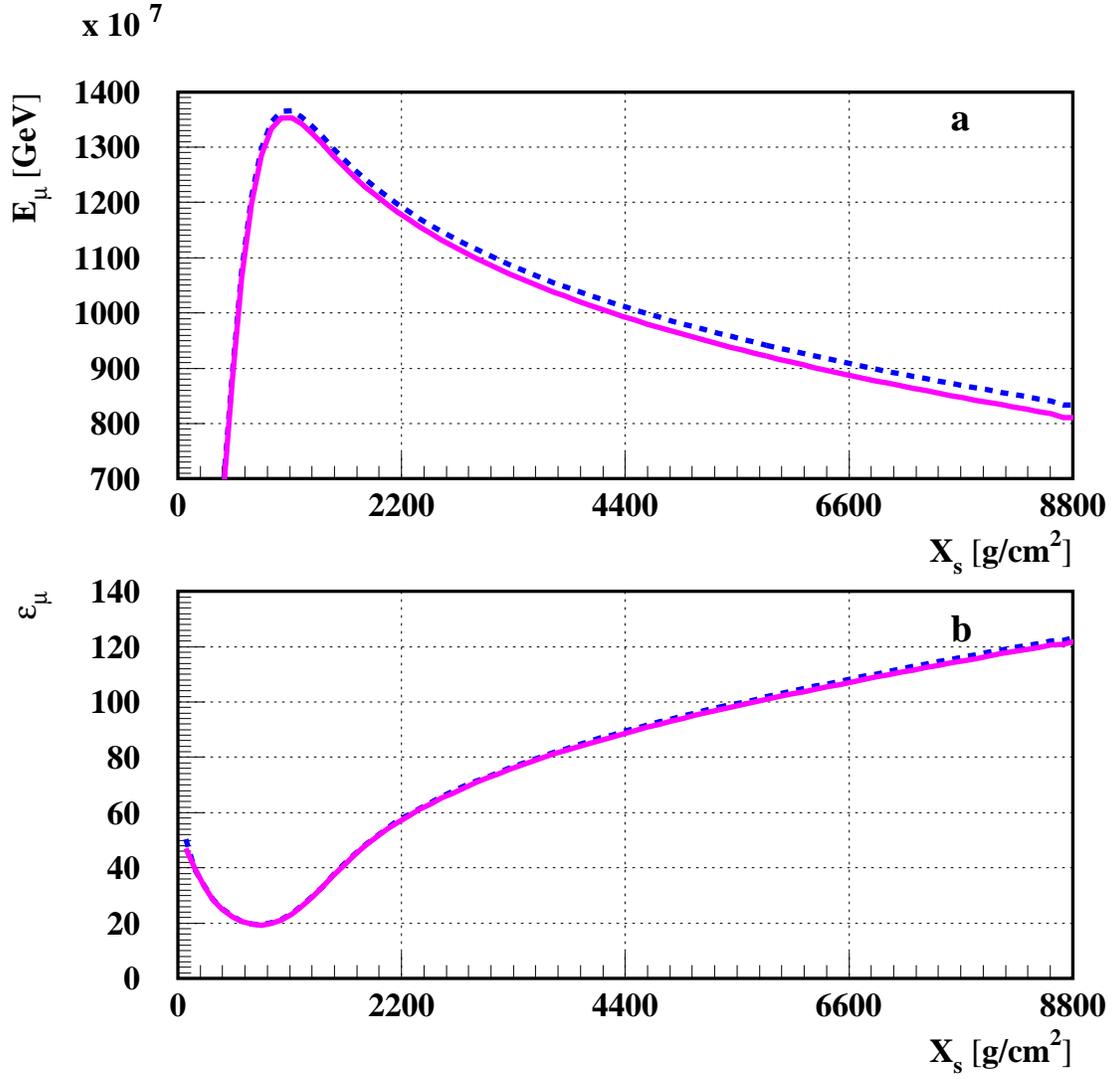}}
\end{displaymath}
\caption{(a) Average longitudinal development  of muon energy  versus
$X_s$. (b) Average energy per muon versus $X_s$.
The solid (dashed) lines corresponds to the case where the MBR and MPP
are  (are not) taken into account. The initial conditions are the same
as in figure \ref{fig:p6_z85_3e20_elctrongammaerrnew}.
}
\label{fig:pldm}
\end{figure}
\clearpage
\begin{figure}
\begin{displaymath}
\hbox{\epsfig{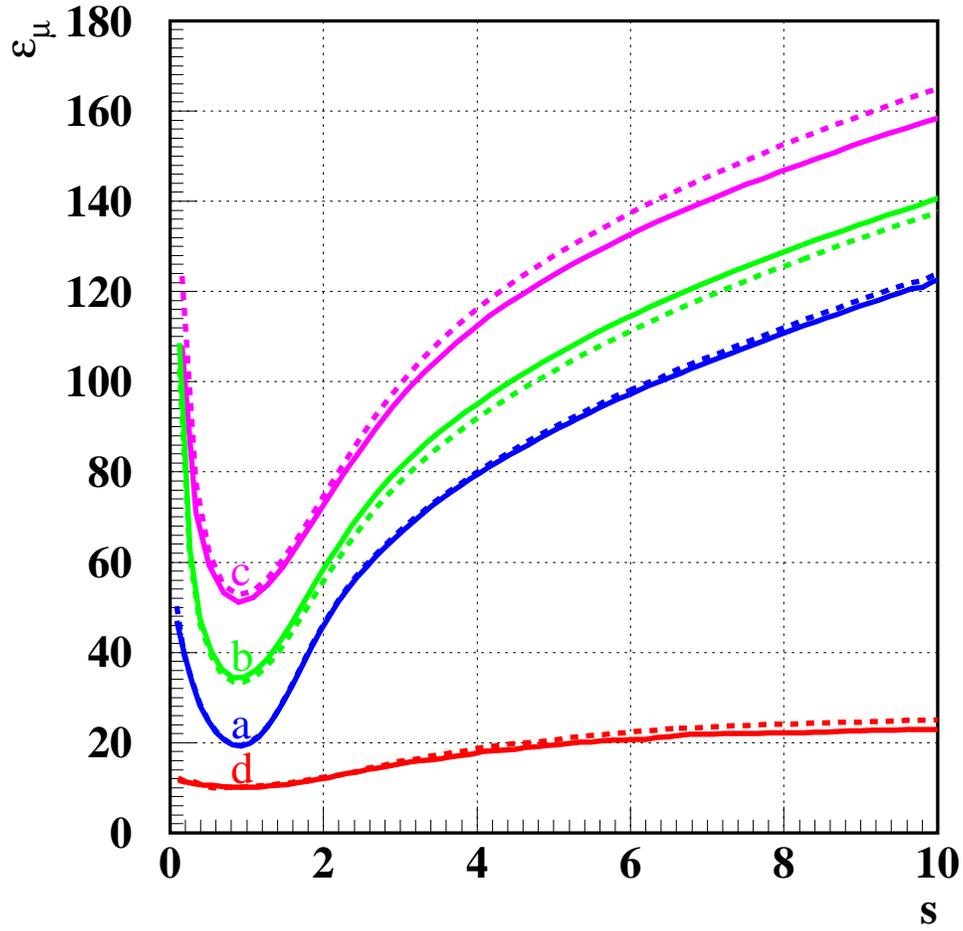}}
\end{displaymath}
\caption{$\xi_{\mu}$ (GeV/particle) versus $s$ (see text), for proton
showers inclined 85 degrees. 
The  solid (dashed) lines correspond to the case where the MBR and MPP
interactions are  (are not) taking into account.
(a), (b), (c) and  (d) correspond, respectively, to primary energies of
$3\times 10^{20}$ eV, $10^{16}$ eV, $10^{14}$ eV and $3\times 10^{11}$
eV. The averages were performed using sets of 200 (a), 250 (b), 500
(c), 1000 (d) showers, in all cases simulated with $10^{-6}$ relative
thinning and weight factor 3, and neglecting the effect of the
geomagnetic field.}
\label{fig:p6_z85vs_snewmuon}
\end{figure}
\clearpage
\begin{figure}
\begin{displaymath}
\hbox{\epsfig{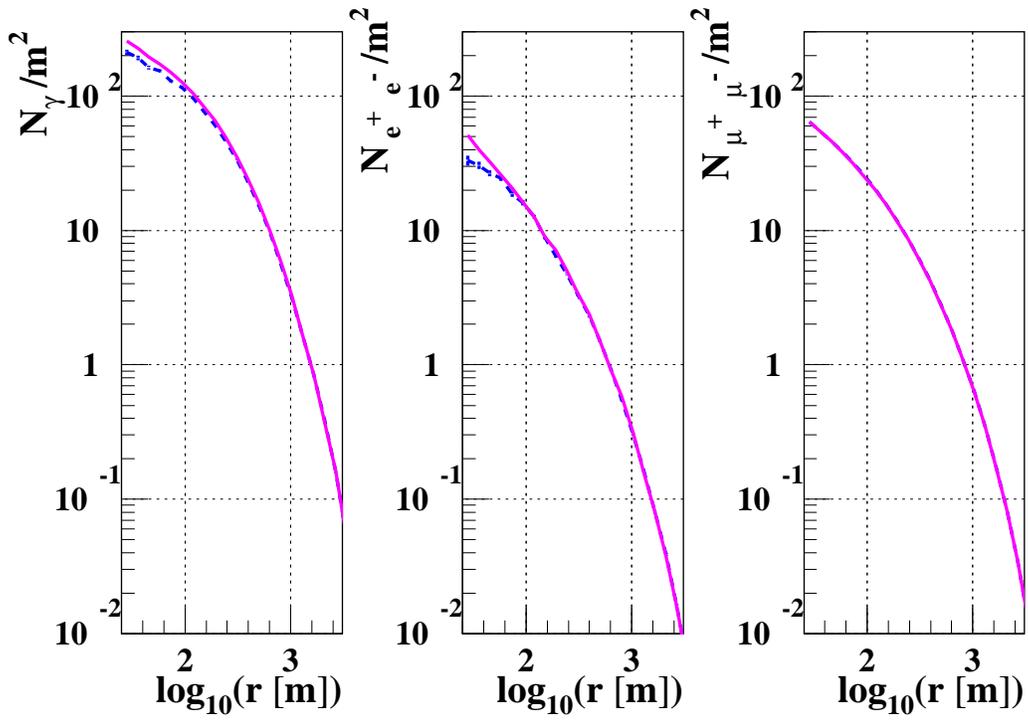}}
\end{displaymath}
\caption{Number of particles ($\gamma$, $e^+e^-$ and $\mu^+\mu^-$
respectively) versus the distance to the shower axis, for $10^{19}$ eV
proton showers with zenith angle $70^{\circ}$.
The solid (dashed) lines correspond to the case where the MBR and MPP
are  (are not) taken into account.
Each curve corresponds to an average over 400 showers simulated with
$10^{-6}$ relative thinning and weight limiting
factor 3. The geomagnetic field was taken into account
during the simulations.
}
\label{fig:lateral}
\end{figure}
\clearpage
\begin{figure}
\begin{displaymath}
\hbox{\epsfig{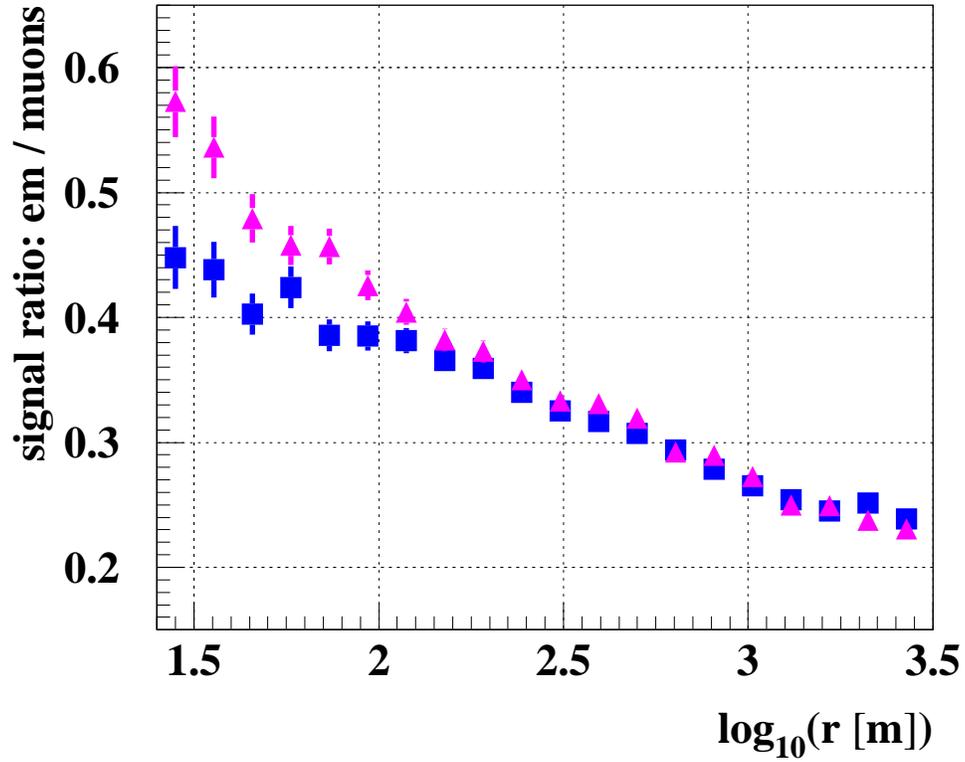}}
\end{displaymath}
\caption{Signal ratio: electromagnetic component divided by muon
component at ground level, plotted versus the distance to the shower
axis. The triangles (squares) correspond to the
case  where the MBR and MPP are (are not) taken into account. The
simulation parameters are the same as in figure \ref{fig:lateral}.}
\label{fig:signal}
\end{figure}

\end{document}